\documentclass[12pt]{article}
\usepackage[dvips]{graphicx}
\usepackage[tight]{subfigure}
\usepackage{amsmath,amssymb,amsfonts}
\usepackage[utf8]{inputenc}

\usepackage[numbers,sort&compress,square]{natbib}
\usepackage{hyperref}
\def\be{\begin{equation}}
\def\ee{\end{equation}}
\def\bea{\begin{eqnarray}}
\def\eea{\end{eqnarray}}
\def\beaN{\begin{eqnarray*}}
\def\eeaN{\end{eqnarray*}}
\def\ed{\end{document}}
\def\bit{\begin{itemize}}
\def\eit{\end{itemize}}

\def\sig{\sigma}

\def\k{\kappa}
\def\alf{\alpha}

\def\di{\partial}

\def\~{\tilde}
\def\lag{{{\cal L}}}
\def\m{\label}
\def\l{\left}
\def\r{\right}

\def\goto{\rightarrow}

\def\diag{\rm diag}

\def\sA{{\stackrel{\bullet}{A}}{}}

\def\cGG{\stackrel{\circ}{G}}
\def\sS{{\stackrel{\bullet}{S}}{}}

\def\sj{\stackrel{\bullet}{j}}

\def\sK{{\stackrel{\bullet}{K}}{}}
\def\sT{{\stackrel{\bullet}{T}}{}}
\def\cT{{\stackrel{\circ}{T}}{}}
\def\sL{\stackrel{\bullet}{\lag}}
\def\cN{\stackrel{\circ}{\nabla}}
\def\cA{{\stackrel{\circ}{A}}{}}
\def\cR{\stackrel{\circ}{R}}

\def\scJ{\stackrel{\bullet}{\cal J}}

\usepackage{authblk}

\usepackage[dvipsnames]{xcolor}
\usepackage{graphicx}

\begin{document}
\title{ \bf \Large ON CONSERVED QUANTITIES FOR THE SCHWARZSCHILD BLACK HOLE IN TELEPARALLEL GRAVITY}
\author[1,2]{E. D. Emtsova\thanks{Electronic address: \texttt{ed.emcova@physics.msu.ru}}}
\author[3,4]{M. Kr\v{s}\v{s}\'ak\thanks{Electronic address: \texttt{martin.krssak@gmail.com}}}
\author[1]{A. N. Petrov\thanks{Electronic address: \texttt{alex.petrov55@gmail.com}}}
\author[1,5]{A. V. Toporensky\thanks{Electronic address: \texttt{atopor@rambler.ru}}}
\affil[1]{Sternberg Astronomical Institute, Lomonosov Moscow State University  \protect\\ Universitetskii pr., 13, Moscow, 119992,
Russia}
\affil[2]{Faculty of Physics, Lomonosov Moscow State University, Moscow 119991, Russia}
\affil[3]{Bangkok Fundamental Physics Group, Department of Physics
 \protect\\ Chulalongkorn University, Bangkok 10330, Thailand}
\affil[4]{Center for Gravitation and Cosmology, College of Physical Science	and Technology  \protect\\
Yangzhou University, Yangzhou 225009, China}
\affil[5]{Kazan Federal University, Kremlevskaya 18, Kazan, 420008, Russia}

\date{\small \today}
\maketitle

\begin{abstract}
We examine various methods of constructing conserved quantities in the Teleparallel Equivalent of General Relativity (TEGR). We demonstrate that in the covariant formulation the preferred method are the Noether charges that are true invariant quantities. The Noether charges depend on the vector field $\xi$ and we consider two different options where $\xi$ is chosen as either a Killing vector or a four-velocity of the observer. We discuss the physical meaning of each choice on the example of the Schwarzschild solution in different frames: static,  freely falling Lemaitre frame, and a newly obtained generalised freely falling frame with an arbitrary initial velocity. We also demonstrate how to determine  the inertial spin connection for various tetrads used in our calculations, and find a certain ambiguity in the ``switching off" gravity method where different tetrads can share the same inertial spin connection.  
\end{abstract}

\section{Introduction}
\setcounter{equation}{0}

Teleparallel theories of gravity became  increasingly popular subject in recent years. This includes the case of reformulation of  general relativity itself, known as the teleparallel equivalent of general relativity (TEGR) \cite{Aldrovandi_Pereira_2013,Maluf,REV_2018}, as well as  various modified gravity models like the  $f(T)$ gravity \cite{Ferraro:2006jd,Bengochea:2008gz,Ferraro:2008ey,Linder:2010py,f(T)3}  and other models \cite{Hayashi:1979qx,Geng:2011aj,Maluf:2011kf,Bahamonde:2017wwk,Hohmann:2017duq}. One of the most attractive features of TEGR is the possibility of constructing various new conserved quantities and address the problem of definition of energy-momentum known to be difficult in the standard general relativity \cite{Szabados_2009,Petrov_KLT_2017}. 

Our primary goal is to clarify the situation with  conserved quantities in teleparallel gravity where two distinct  approaches to their definition exist. The first approach follows from direct integration of the field equations that yield the quantity $P_a$  identified as the total energy-momentum \cite{Maluf0704,Obukhov+,Aldrovandi_Pereira_2013}. The second approach is based on the application of the Noether theorem that lead to the Noether conserved charges  \cite{Obukhov:2006ge,Obukhov_Rubilar_Pereira_2006}.  The total energy-momentum $P_a$ is not an invariant quantity, and there exists, at least in principle, an ambiguity whether  to consider its Lorentz or spacetime indexed version. The  Noether conserved charges $P(\xi)$, on the other hand, are true invariant quantities but depend on the additional vector field $\xi$ characterizing the observer that need to be specified.

We demonstrate that the first approach can be understood as a special case of the Noether approach  where the vector field $\xi$ is chosen to  coincide with a certain  tetrad vector and obtain the physical interpretation of this result.
However, to our surprise, we find that $P_a$  defines the energy-momentum uniquely only in the case of the non-covariant formulation of TEGR where the spin connection is assumed to be zero. This follows from the fact that in the non-covariant formulation, we can uniquely identify the vector field $\xi$ characterizing the observer with the tetrad variable in the field equations.  In the covariant formulation, on the other hand, the freedom to use an arbitrary tetrad in the field equations forces us to treat the vector field $\xi$ independently and hence  use the Noether conserved charges in order to obtain meaningful  results.

We apply both approaches on the example of the Schwarzschild solution, where we consider both the static and the free-falling observers.  This follows the logic of previous works by Maluf \textit{et. al.} \cite{Maluf0704} and Lucas \textit{et. al.} \cite{Obukhov+}, where the conserved quantities for the free-falling observer were considered. However, both \cite{Maluf0704} and  \cite{Obukhov+}, calculated only the zeroth component of $P_a$ identified as the energy, in the non-covariant and covariant formulations, respectively. The Noether conserved charges, on the other hand, were introduced  in works of Obukhov \textit{et. al.}  \cite{Obukhov:2006ge,Obukhov_Rubilar_Pereira_2006,Obukhov_2006}, but, in the case of the Schwarzschild solution, these were calculated only in the case of the static observer.  We argue that  while in the static case both notions of the conserved quantities coincide, the free-falling case clearly reveals  important differences.

Our second goal is to illustrate the problem of determining the inertial spin connection in the covariant formulation of teleparallel gravity that we encounter in our calculations. In the covariant formulation, both the tetrad and  inertial spin connection are the  fundamental variables. While the field equations determine the tetrad up to a local Lorentz transformation, the purely inertial spin connection is not determined by any field equations. Nevertheless, the choice of a spin connection and the local Lorentz degrees of freedom in the tetrad do influence some physically relevant quantities, including the conserved charges that we are interested in here.

The idea of the covariant  formulation is that the inertial spin connection must be chosen according to the tetrad to remove divergences of the relevant physical quantities. This was first introduced as an ansatz in  \cite{Obukhov:2002tm} and later it was argued to be equivalent to \textit{switching-off} gravity by taking the $r\rightarrow\infty$ limit of the Levi-Civita connection corresponding to the tetrad ansatz \cite{Obukhov:2006ge,Obukhov_Rubilar_Pereira_2006,Obukhov_2006,Obukhov+}. It was then advocated that the underlying principle can be altered  to choose the spin connection in a such way that  the teleparallel Lagrangian vanishes asymptotically \cite{Krssak:2015rqa,Krssak:2015lba,REV_2018}. In \cite{EPT19,EPT_2020}, the idea of ``switching of'' gravity was further formalized in order to be less dependent  on the asymptotic behaviour. Nevertheless, at least for the asymptotically Minkowski spacetimes, all these various approaches reduce to the original principle of \textit{switching-off} gravity, and the inertial spin connection is determined from the Levi-Civita spin connection by  taking a parameter that controls the strength of gravity to zero or considering the limit $r\rightarrow\infty$.

However, we encounter a problem with this approach when we try to apply it in the case of both the static and free-falling observers. Our starting points are spacetime metrics for the Schwarzschild solution written in both the spherical  and  Lemaitre coordinate systems. In each case,  we choose the diagonal tetrad and determine the spin connection by switching-off gravity. Despite these tetrads representing two different physical observers, i.e. the static one and the free-falling one, we obtain the same expression for the inertial spin connection. This demonstrates an ambiguity of the above procedure and we argue that it is related to the problem of the so-called remnant symmetries discovered initially in $f(T)$ gravity \cite{Ferraro:2014owa,Chen:2014qtl,Krssaktbp}.

Our third result is a generalization of the situation with free-falling observers to the case of  observers with an arbitrary initial velocity. This  leads us to define a new gauge that we name the generalized Lemaitre gauge or $e$-gauge. We demonstrate that in the case when the vector field $\xi$ is identified with the observer velocity, we find the vanishing Noether current and charge, analogously to the ordinary Lemaitre gauge. However, in the case when we identify the vector field $\xi$ with the time-like Killing field, we find curiously the same Noether charge as in the Schwarzschild static gauge. This is in contrast with the ordinary Lemaitre gauge where an analogous calculation leads to an unphysical result.

As the last point, we consider the analogous situation in the most popular modified teleparallel model known as the $f(T)$ gravity model \cite{Ferraro:2006jd,Bengochea:2008gz,Ferraro:2008ey,Linder:2010py,f(T)3}. Here we face a similar situation as in TEGR since we have to determine the spin connection corresponding to the tetrads in the covariant formulation \cite{Krssak:2015oua,REV_2018}, or find the special class of tetrads, nick-named as \textit{good tetrads}, with corresponding vanishing inertial spin connections \cite{Tamanini:2012hg}. The problem of determining the spin connection is then even more pressing  than in TEGR since it affects the field equations and hence becomes the problem of dynamics. We show that the analogous construction of the Lemaitre gauge does not work for a general spherically symmetric spacetime as it does not yield a new good tetrad. However, we show  that it is possible to find a specific boost that makes the torsion scalar vanish. While this does not lead to new $f(T)$ gravity solutions since the theory reduces to TEGR, these kind of solutions are interesting since these vanishing torsion scalar solutions can be used to demonstrate that  GR solutions stay solutions in the $f(T)$ gravity as well \cite{Ferraro:2011ks,Bejarano:2014bca,Bejarano:2017akj}.

The paper is organized as follows. In section~\ref{secrev}, we briefly review some basic elements of teleparallel gravity. In section~\ref{seccharges}, we introduce various conserved currents introduced in the literature previously and we discuss their relations.

In section~\ref{3}, we apply these definitions to the Schwarzschild case, where our starting point is the Schwarzschild metric in the spherical coordinate system and we choose the static tetrad. We illustrate the principle of determining the inertial spin connection and find the so-called proper tetrad,  and  introduce a notion of the Schwarzschild static gauge, where different combinations of tetrads and corresponding spin connections represent the same physical situation. We calculate the conserved charges and obtain the physically expected result for the Noether charge that can be identified with the total energy of the black hole.

In section~\ref{4}, we follow the same logic and calculate the same conserved quantities but we start with the metric written in the free-falling Lemaitre coordinates and the free-falling tetrad. This leads us to define in the analogous fashion the so-called Lemaitre gauge.  We show that we  obtain naturally the zero energy density measured by the freely falling observer in this case.

In section~\ref{5}, we express all quantities from the Lemaitre  gauge in the standard static Schwarzschild coordinates. Then we compare them with  quantities in the Schwarzschild static  gauge and analyse how the conserved charges depend on the gauge. On the one hand, in the framework of the Schwarzschild static gauge we obtain the correct total mass, but we have a problem with a description of the freely falling observer. On the other hand, we demonstrate that while the Lemaitre gauge offers a correct description of the freely falling observer,  if we try to recover calculation of the total energy by choosing  $\xi$ to be the time-like Killing vector, we obtain a problematic result. We provide the detailed comparison of our results to the previous authors as well.

In section~\ref{secegauge}, we generalize the Lemaitre gauge by considering an arbitrary initial velocity, leading us to the generalized Lemaitre gauge that we call the $e$-gauge. We demonstrate that the generalized the $e$-gauge leads to either vanishing conserved currents and charges or the result corresponding to the total black hole mass, depending on the choice of the field $\xi$. These results are compared with the result from the previous Lemaitre gauge.

In section~\ref{secfT}, we consider the accelerated tetrads  in  $f(T)$ gravity. We demonstrate that the analogue of the Lemaitre gauge does not lead to any new good tetrad and we show how to construct a solution with a vanishing torsion scalar.

In section~\ref{secdiscfin}, we summarize our results and draw some final conclusions. All the lengthy expressions for the inertial  spin connections corresponding to various tetrads are summarized in Appendix~\ref{appendA}.

\textit{Notation:} We mostly follow here the notation used in  the book \cite{Aldrovandi_Pereira_2013}, but we work in the mostly positive convention $(-,+,+,+)$.  By Latin indices we denote the tangent space coordinates and Greek indices represent the spacetime coordinates. When we work with the particular components we distinguish tangent indices by a hat, e.g. in $\cA{}^{\hat{1}}_{\ \hat{2}1}$ two first indices are tangent space components and $1$ is a spacetime component. Teleparallel quantities are denoted with ``$\bullet$"  over them and those with ``$\circ$"  are related to the Levi-Civita connection. For example, $\sT^a{}_{\mu\nu}$ and  $\cT^a{}_{\mu\nu}$, represent the torsion tensors of the teleparallel and Levi-Civita connections, respectively. For the sake of simplicity, we do omit ``$\bullet$"  when we make reference to $f(T)$ gravity in the text, although according to our notation it should be $f(\sT)$ gravity.

\section{Overview of teleparallel gravity\label{secrev}}
\setcounter{equation}{0}

In this section, we briefly introduce the teleparallel equivalent of general relativity (TEGR).
The Lagrangian of TEGR is given by \cite{Aldrovandi_Pereira_2013}
\be
\sL =  
\frac{h}{2\kappa}\sT\,
\equiv
\frac{h}{2\kappa} \l(\frac{1}{4} {\sT}{}^\rho{}_{\mu\nu} {\sT}_\rho{}^{\mu\nu} + \frac{1}{2} {\sT}{}^\rho{}_{\mu\nu} {\sT}{}^{\nu\mu}{}_\rho - {\sT}{}^\rho{}_{\mu\rho} {\sT}{}^{\nu\mu}{}_\nu\r),
\m{lag}
\ee
where $\sT$ is the torsion scalar and ${\sT}{}^a{}_{\mu\nu}$ is the torsion tensor defined as
\begin{equation}\label{tor}
{\sT}{}^a{}_{\mu\nu} = \di_\mu h^a{}_\nu - \di_\nu h^a{}_\mu + {\sA}{}^a{}_{c\mu}h^c{}_\nu - {\sA}{}^a{}_{c\nu}h^c{}_\mu,
\end{equation}
with the tetrad components $h^a{}_\nu$, $h = \det h^a{}_\nu$, $\kappa=8 \pi G$ (in $c=1$ units), and the inertial spin connection ${\sA}{}^a{}_{c\nu}$ given by
\begin{equation}
 {\sA}{}^a{}_{c\nu}=\Lambda^a {}_b \partial_\nu    (\Lambda^{-1}){}^b {}_c,
\label{telcon}
\end{equation}
where $\Lambda^a_{\ c}$ is a matrix of a local Lorentz transformation.

We can define the contortion tensor 
\begin{equation}\label{tor_K}
    \sK^\rho{}_{\mu\nu}=\frac{1}{2}(\sT_\mu{}^\rho{}_\nu+\sT_\nu{}^\rho{}_\mu -\sT^\rho{}_{\mu\nu})
\end{equation}
that directly relates the inertial spin connection \eqref{telcon}  with the  Levi-Civita spin connection of general relativity
\be
\sK^a{}_{b\rho} = \sA^a{}_{b\rho} - \cA^a{}_{b\rho},
\m{K_A_A}
\ee
where $\cA^a{}_{b\rho} $ is the usual  Levi-Civita spin connection  defined by
\begin{equation}\label{A}
    \cA{}^a{}_{b\mu} = -h_b{}^\nu \cN_\mu h^a{}_\nu.
\end{equation}
In addition, we  define the superpotential
\begin{equation}\label{super_K}
     {\sS}_a{}^{\rho\sigma}=
     %- \frac{\kappa}{h} \frac{\di \sL}{\di h^a{}_{\rho,\sigma}}
     \sK{}^{\rho\sigma} {}_a + h_a{}^{\sigma} \sK{}^{\theta \rho} {}_{\theta} - h_a{}^{\rho} \sK{}^{\theta \sigma} {}_{\theta}\, ,
\end{equation}
which is antisymmetric in the last two indices and is used to define the torsion scalar in \eqref{lag} as
\begin{equation}\label{torscalar}
\sT=\frac{1}{2}\,{\sS}{}_a{}^{\rho\sigma}{\sT}{}^a{}_{\rho\sigma}.
\end{equation}

All tensors ${\sT}{}^a{}_{\mu\nu}$, $\sK^{\rho\sigma}{}_{a}$, and ${\sS}_a{}^{\rho\sigma}$, are true tensors with respect to both coordinate transformations and local Lorentz transformations, while the torsion scalar $\sT$ is a scalar. Note that working in the covariant formulation of teleparallel gravity, the local Lorentz covariance means that the tensorial quantities transform covariantly under the simultaneous transformation of both the tetrad $h'{}^a {}_{\mu} = \Lambda {}^a {}_b  h^b{}_\mu$ and the inertial spin connection
\begin{equation}\label{spin_trans}
\sA{}'{}^a {}_{b \mu}=\Lambda {}^a {}_c \sA{} {}^c {}_{d \mu} \Lambda {}_b {}^d   + \Lambda {}^a {}_c \partial_\mu \Lambda {}_b {}^c ,
\end{equation}

The inertial spin connection $\sA{}^a{}_{b \rho}$ is not determined by the field equations but, as we have discussed in Introduction, it can be chosen by the principle of "switching-off"  gravity \cite{ Obukhov+,Krssak:2015rqa,Krssak:2015lba,REV_2018}. According to \cite{EPT19,EPT_2020}, this means that for the tetrad  $h^a{}_\mu$ we calculate the Levi-Civita connection (\ref{A}) and the Riemann tensor $\cR{}^a{}_{b\mu\nu}$. Then, we find parameters which allow us to continuously ``switch-off" gravity, i.e. to obtain
\begin{equation}\label{B}
   \cR {}^a{}_{b\mu \nu}=0.
\end{equation}
In practice, for the asymptotically Minkowski spacetimes, this means that we consider either $r\rightarrow\infty$  or $M\rightarrow0$ limit of the Levi-Civita connection that leads to \eqref{B}.

Applying the variational principle one  obtains the field equations in the form
\begin{equation}\label{FE}
E_a{}^\rho=h \kappa \Theta_a{}^\rho,
\end{equation}
where on the LHS we have defined the Euler-Lagrange expression
 \be
E_a{}^\rho\equiv \frac{\di \sL}{\di h^a{}_\rho} - \di_\sig \l(\frac{\di \sL}{\di h^a{}_{\rho,\sigma}} \r)
=
\partial_\sigma\Big(h\sS_a{}^{\rho \sigma}\Big) -
\kappa \, h \sj_{a}{}^{\rho},
 \m{EM+}
 \ee
where
 \begin{equation}
 \sj_{a}{}^{\rho}= \frac{1}{\k}h_a{}^\mu \sS_c{}^{\nu\rho}\sT^c{}_{\nu\mu} -\frac{h_a{}^\rho}{h}\sL + \frac{1}{\k}\sA^c{}_{a\sig}\sS_c{}^{\rho\sig},
\end{equation}
is the gravitational energy-momentum current, and on the RHS we have defined the matter energy-momentum tensor corresponding to the matter Lagrangian $\lag_m$ as
\be
\Theta_a{}^\rho = -\frac{1}{h}\frac{\delta \lag_m}{\delta h^a{}_\rho}\,.
\m{matterEM}
\ee

\section{Total energy-momentum and Noether conserved \\charges\label{seccharges}}
\setcounter{equation}{0}
The central point of this paper is calculation of the conserved quantities in TEGR, and as we have mentioned in Introduction, there are two distinct  definitions of the conserved charges that we will  introduce now and discuss in detail.

The first definition is based on the field equations \eqref{FE}, since, as we can observe, they naturally define  the conserved quantities by taking an ordinary derivative
\begin{equation}
\partial_\rho (h \sj_a{}^\rho+h \Theta_a{}^\rho)=0,
\end{equation}
on the account of the superpotential being antisymmetric in the last two indices and partial derivatives commuting. The term in the bracket defines a conserved quantity
\begin{equation}
P_a=-\int_\Sigma d^3x (h \sj_a{}^\rho+ h \Theta_a{}^\rho).
\end{equation}
where the sign is chosen to produce a positive $P_{\hat{0}}$ in order to be identified with the energy.

Using the Stokes theorem and the field equations \eqref{FE}, it can be written in the standard spherical coordinates as
\begin{equation}
P_a =-\lim_{r\rightarrow\infty}\frac{1}{\kappa}
\int_{\partial \Sigma}d^2x \, h \sS_a^{\  01}.  \label{Pa}
\end{equation}
This is the definition  discussed in the book \cite{Aldrovandi_Pereira_2013}, or references \cite{Maluf0704} or \cite{Obukhov+}, where it is identified as the total energy-momentum, and hence in the absence of  matter as the gravitational energy-momentum.

However, there are two subtle points that we would like to highlight here. The first one is that  it behaves as a vector under local Lorentz transformations. As it was noted in \cite{Obukhov_2006}, this quantity is invariant under the  global Lorentz transformations and local Lorentz transformations that vanish at the infinity. However, this is not enough to make it a true invariant conserved charge. 

The second point is that there is an ambiguity in this definition that stems from which form of the field equations we start with. We can write the field equations \eqref{FE} in a fully spacetime form, and, analogously to the procedure above, define the spacetime indexed total energy-momentum as \cite{Krssak:2015rqa}
\begin{equation}
P_\mu =-\lim_{r\rightarrow\infty}\frac{1}{\kappa}
\int_{\partial \Sigma}d^2x \, h \sS_\mu^{\  01}. \label{Pmu}
\end{equation}
While being superficially similar to \eqref{Pa}, this quantity is  different. Most obviously, it is invariant with respect to all local Lorentz transformations (in our covariant formulation), but not diffeomorphism invariant since it transforms as a spacetime vector.

We will demonstrate that both of these points can be understood by considering the second definition based on the diffeomorphism invariance and the Noether theorem. This is the definition first introduced by Obukhov \textit{et. al.} \cite{Obukhov:2006ge,Obukhov_2006,Obukhov_Rubilar_Pereira_2006} in the language of differential forms and derived in a more common tensorial language  in the recent papers \cite{EPT19,EPT_2020}. Here we will only briefly outline the main idea, while all the details can be find in \cite{EPT19,EPT_2020}.

The Noether definition of the energy-momentum follows from considering the invariance under diffeomorphisms induced by an arbitrary vector field $\xi$. This naturally leads to the Noether current
${\scJ}{}^\alf(\xi)$ defined as
\begin{equation}
{\scJ}{}^\alf(\xi)=\partial_\beta {\scJ}{}^{\alf\beta}(\xi),
\end{equation}
where 
\begin{equation}\label{super}
 {\scJ}{}^{\alf\beta}(\xi)=
  \frac{h}{\kappa}\xi^\sigma \sS_\sigma{}^{\alpha\beta},
\end{equation}
is the Noether current superpotential.

By  construction, the current ${\scJ}{}^{\alf}(\xi)$ is a vector density, the superpotential ${\scJ}{}^{\alf\beta}(\xi)$ is an antisymmetric tensor density, and both ${\scJ}{}^{\alf}(\xi)$ and ${\scJ}{}^{\alf\beta}(\xi)$ are locally Lorentz invariant. The Noether current is conserved in both the ordinary and covariant sense, i.e.
\be
\di_\alf {\scJ}{}^\alf(\xi) = \cN_\alf {\scJ}{}^\alf(\xi) =0,
\m{DiffCL_2}
\ee
where $\cN_\alf$ is the Levi-Civita covariant derivative, what allows us to define the Noether conserved charge as
\begin{equation}\label{noetcharge}
    {\cal P}(\xi) = \int_\Sigma d^3x  {\scJ}{}^0(\xi) = \oint_{\di\Sigma} d^2 x{\scJ}{}^{01}(\xi)\,,
\end{equation}
which can be written in the spherical coordinate system explicitly as
\begin{equation}\label{netercharge}
    {\cal P}(\xi) =   \lim_{r\rightarrow\infty}\frac{1}{\kappa}
\int_{\partial \Sigma}d^2x \, h\, \xi^\sigma \sS_\sigma^{\  01}.
\end{equation}

In particular case when $\xi$ is time-like, we would like to identify the corresponding conserved charge with the energy measured by a set of observers at spatial infinity with their four-vectors $\xi$ 
\begin{equation}\label{Energy}
E(\xi)={\cal P}(\xi), \qquad \xi \!: \text{time-like}.
\end{equation}

Unlike the previous quantities \eqref{Pa} and \eqref{Pmu},  the Noether charges \eqref{netercharge} are  true invariant quantities with respect to both diffeomorphism and local Lorentz transformations.

The price for this invariance  is that the Noether charges \eqref{netercharge} depend on the vector field $\xi$ that is not  fixed \textit{a priori} and needs to be determined.  In principle, we have many possible ways to choose $\xi$, but naturally some of these choices are more  physically meaningful. In the standard metric formulation of GR, it was argued that the physically preferred choice is to choose  $\xi$ to be a  Killing field of the reference geometry \cite{Chen:1998aw}.

However,  in the case of a tetrad theory, as is the case of TEGR, we have additional freedom due to the fact that the tetrad is not fully determined by the field equations. This naturally provides us with another possibility to choose  $\xi$ to be identified with the tetrad itself that allows us to understand the energy-momentum \eqref{Pa} within the context of the Noether charges. To illustrate this, let  us first identify  the time-like vector $\xi^\mu$  with the zeroth tetrad vector, i.e. $\xi^\mu=-h_{\hat{0}}{}^\mu$, which  naturally leads to $\mathcal{P}(\xi)=P_{\hat{0}}$. We can then generalize this to four vector fields labelled as $\xi_{(a)}$ and identify them with the tetrad as $\xi_{(a)}=h_a{}^\mu$ to obtain
\begin{equation}
\mathcal{P}(\xi_{(a)})=P_{a}.
\end{equation}
Therefore, the total energy-momentum $P_a$ defined by \eqref{Pa} indeed represents a physically interesting quantity, i.e. these are the Noether charges for the diffeomorphisms generated by the tetrad vectors themselves. We can then understand $P_a$  as the total energy-momentum of the whole spacetime as measured by the observer represented by the tetrad.

It is interesting to observe that applying the  same logic to \eqref{Pmu}, we find that $P_\mu$ are the Noether charges for $\xi^\mu=dx^\mu$, i.e. diffeomorphisms generated by the coordinates. This makes the physical meaning of $P_\mu$ rather unclear, unless one provides a further information how to attach some meaning to the coordinates. One could in principle argue that $P_\mu$ represents the energy-momentum of the spacetime itself since the quantity is Lorentz scalar in the covariant formulation. However, as we demonstrate shortly, the problem is that there exist various gauges and hence the superpotential can have different values for different observers.

\section{Black hole mass and Schwarzschild static gauge}
\setcounter{equation}{0}
\m{3}

Let us demonstrate calculation of some of these quantities related to determining the mass of the Schwarzschild black hole.
The standard metric for the Schwarzschild black hole is
\begin{equation}
    ds^2=-fdt^2+f^{-1}dr^2+r^2 (d\theta^2 + \sin^2\theta d\varphi^2),
    \label{BHmet}
\end{equation}
where
\begin{equation}\label{f(r)}
    f=f(r) = 1-\frac{2M}{r}.
\end{equation}
The tetrad is not defined by the field equations \cite{Aldrovandi_Pereira_2013}, so, we can choose the most convenient diagonal tetrad as
\begin{equation} \label{BHtet}
{\stackrel{B}{h}}{}^{a} {}_{\mu } \equiv {\diag}{\left(f^\frac{1}{2}, f^{-\frac{1}{2}}, r, r\sin \theta \right)}\, .
\end{equation}

The Levi-Civita spin connection for the tetrad (\ref{BHtet}), defined by (\ref{A}), has non-zero components
\begin{gather}\label{BHGRspin}
\cA^{\hat{0}} {}_{\hat{1}0}=\cA^{\hat{1}} {}_{\hat{0} 0}=\frac{M}{r^{2} },\qquad
\cA^{\hat{1}} {}_{\hat{2}2}=-\cA^{\hat{2}} {}_{\hat{1}2} =-f^{\frac{1}{2}},\\
\cA^{\hat{1}} {}_{\hat{3}3}=-\cA^{\hat{3}} {}_{\hat{1}3}=-{f^\frac{1}{2}\sin\theta ,}\qquad
\cA^{\hat{2}} {}_{\hat{3}3}=-\cA^{\hat{3}} {}_{\hat{2}3}=-\cos\theta.
\nonumber
\end{gather}

We find that all components of the Riemannian curvature are proportional to $M$ and hence the limit $M\goto 0$ of (\ref{BHGRspin}) defines the inertial spin connection $\stackrel{\bullet}{A} {}^{a} {}_{c\mu }$ related to  (\ref{BHtet}), non-zero components of which are
\begin{equation} \label{BHspin}
\stackrel{\bullet}{A}{}^{\hat{1}} {}_{\hat{2}2} =-\stackrel{\bullet}{A} {}^{\hat{2}} {}_{\hat{1}2} =-1,\quad  \stackrel{\bullet}{A} {}^{\hat{1}} {}_{\hat{3}3} =-\stackrel{\bullet}{A} {}^{\hat{3}} {}_{\hat{1}3} =-\sin \theta,\qquad \stackrel{\bullet}{A} {}^{\hat{2}} {}_{\hat{3}3} =-\stackrel{\bullet}{A} {}^{\hat{3}} {}_{\hat{2}3} =-\cos \theta .
\end{equation}
Using \eqref{super_K} we find that the non-vanishing components of the superpotential are
\begin{equation}\label{superSG}
    \begin{array}{cccc}
   \sS{}_{0} {}^{0 1} = - \sS{}_{0} {}^{1 0} =
   \frac{2}{r} \l( f -f^{\frac{1}{2}} \r),\quad
\sS{}_{2} {}^{1 2} = \sS{}_{3} {}^{1 3} = - \sS{}_{2} {}^{2 1} = - \sS{}_{3} {}^{3 1} =-\frac{1}{r} \l((1+f)/2 -f^{\frac{1}{2}} \right) .\\
    \end{array}
\end{equation}
In the limit $r\rightarrow\infty$, the only non-vanishing components of the corresponding superpotential density are given by
\begin{equation}\label{asymptSchw}
\lim_{r\rightarrow\infty} h \sS{}_{0} {}^{0 1} = - \lim_{r\rightarrow\infty} h \sS{}_{0} {}^{1 0} = -2M \sin\theta.
\end{equation}

Then, choosing  the $\xi$ vector to be the  time-like Killing vector
\begin{equation}\label{Killingtime}
\xi^\alpha = (-1,~0,~0,~0),
\end{equation}
 and using the expressions (\ref{netercharge}), \eqref{Energy}, and (\ref{super}),  we find the Noether charge to be equal to the total mass of the black hole
\begin{equation}\label{E=M}
\mathcal{P}(\xi)=M.
\end{equation}

Applying the Lorentz rotation
\begin{equation} \label{GrindEQ__36_}
(\Lambda_{(Sch)}^{-1}) ^{a} {}_{b} =\left[\begin{array}{cccc} {1} & {0} & {0} & {0} \\ {0} & {\sin \theta \cos \varphi } & {\cos \theta \cos \varphi } & {-\sin \varphi } \\ {0} & {\sin \theta \sin \varphi } & {\cos \theta \sin \varphi } & {\cos \varphi } \\ {0} & {\cos \theta } & {-\sin \theta } & {0} \end{array}\right]
\end{equation}
 to diagonal tetrad  (\ref{BHtet}) one finds that the tetrad components become
\begin{equation} \label{GrindEQ__37_}
{\stackrel{A}{h}}{}^{a} {}_{\mu } \equiv \left[\begin{array}{cccc} f^\frac{1}{2}  & {0} & {0} & {0} \\ {0} & {f^{-\frac{1}{2}} \sin \theta \cos\varphi } & {r\cos \theta \cos \varphi } & {-r\sin \theta \sin \varphi } \\ {0} & {f^{-\frac{1}{2}}\sin \theta \sin \varphi } & {r\cos \theta \sin \varphi } & {r\sin \theta \cos \varphi } \\ {0} & {f^{-\frac{1}{2}}\cos\theta} & {-r\sin \theta } & {0} \end{array}\right]
\end{equation}
and the inertial spin connection vanish. Such a tetrad is called as a {\it proper} tetrad, see \cite{Obukhov+}.

We denote the static diagonal tetrad (\ref{BHtet})  and the related inertial spin connection (\ref{BHspin}) as the  {\em Schwarzschild static gauge}.  Any tetrad with the related inertial spin connection obtained from (\ref{BHtet}) and (\ref{BHspin}) by arbitrary coordinate transformations or arbitrary  local Lorentz rotations \cite{Aldrovandi_Pereira_2013} also represent the Schwarzschild static gauge. Therefore, the proper tetrad (\ref{GrindEQ__37_}) with vanishing inertial spin connection is in the Schwarzschild static gauge  as well.

\section{Free-falling observers and Lemaitre gauge}
\setcounter{equation}{0}
\m{4}

In the present section we study conserved charges in the frame of a freely and radially falling observer into the Schwarzschild black hole. It is natural and reasonable to start the study in the coordinates of the freely falling observer. The most known and popular such coordinates are the Lemaitre ones \cite{Landau_Lifshitz_1975}.

After applying the coordinate transformations from the Schwarzschild static coordinates $(t,r,\theta, \varphi)$ to Lemaitre freely falling coordinates $(\tau,\rho,\theta, \varphi)$
\begin{equation}\label{SchtoLemTransf}
\begin{array}{cccc}
    d\rho =dt+ \frac{dr}{f\sqrt{1-f}}\,,  \\
d\tau =dt+\frac{dr}{f}\sqrt{1-f},
\end{array}
\end{equation}%
with $f(r)$ is defined in (\ref{f(r)}), the  Schwarzschild metric (\ref{BHmet}) transforms to
\begin{equation} \label{metriclem}
ds^2=-d \tau^2+(1-f)d \rho^2+r^2 d\theta^2 + r^2 \sin^2 \theta d\varphi^2 ,
\end{equation}
where
$$r=r(\tau, \rho)=[\frac{3}{2} (\rho-\tau)]^{2/3} (2M)^{1/3}.$$
Now, it is convenient to take the diagonal tetrad related to the metric (\ref{metriclem}):
\begin{equation}\label{Lemtet}
    {\stackrel{C}{h}}{}^a{}_\mu={\diag} \l(1,\sqrt{1-f(r(\tau, \rho))}, r(\tau, \rho), r(\tau, \rho) \sin \theta \r)\,.
\end{equation}

The non-zero Levi-Civita spin connection components  (\ref{A}), corresponding to the tetrad (\ref{Lemtet}), are given by
\begin{gather}
\!\!\!\!\cA{}^{\hat{0}}{}_{\hat{1}1}=\cA{}^{\hat{1}}{}_{\hat{0}1}=\frac{1}{3 (\rho -\tau)} \l(\frac{\frac{4}{3} {M}}{\rho -\tau }\r)^{\frac{1}{3}}\!\!\!\!,\:
\cA{}^{\hat{0}}{}_{\hat{2}2}=\cA{}^{\hat{2}}{}_{\hat{0}2}=
-\l(\frac{\frac{4}{3} {M}}{\rho -\tau }\r)^{\frac{1}{3}}\!\!\!\!,\:
\cA{}^{\hat{0}}{}_{\hat{3}3}=\cA{}^{\hat{3}}{}_{\hat{0}3}=
-\l(\frac{\frac{4}{3} {M}}{\rho -\tau }\r)^{\frac{1}{3}}\!\!\!\!\sin\theta,\nonumber
\\
\cA{}^{\hat{1}}{}_{\hat{2}2}=-\cA{}^{\hat{2}}{}_{\hat{1}2}=-1,\quad
\cA{}^{1}{}_{\hat{3}3}=-\cA{}^{\hat{3}}{}_{\hat{1}3}=-\sin\theta,\quad
\cA{}^{\hat{2}}{}_{\hat{3}3}=-\cA{}^{\hat{3}}{}_{\hat{2}3}=-\cos\theta.
\label{LemLevi}
\end{gather}
All components of the Riemann tensor $\cR{}^a{}_{b \mu \nu}$ calculated from (\ref{LemLevi}) are proportional to $M$ and hence vanish in the limit $M\goto 0$. Then the inertial spin connection $\sA{} {}^a{}_{b \mu}$ can be taken as the $M\goto 0$ limit of the  Lemaitre Levi-Civita spin connection (\ref{LemLevi}), i.e.
\begin{equation}\label{wmlem}
   \stackrel{\bullet}{A} {}^{\hat{1}} {}_{\hat{2}2} =-\stackrel{\bullet}{A} {}^{\hat{2}} {}_{\hat{1}2} =-1, \quad \stackrel{\bullet}{A} {}^{\hat{1}} {}_{\hat{3}3} =-\stackrel{\bullet}{A} {}^{\hat{3}} {}_{\hat{1}3} =-\sin \theta, \quad \stackrel{\bullet}{A} {}^{\hat{2}} {}_{\hat{3}3} =-\stackrel{\bullet}{A} {}^{\hat{3}} {}_{\hat{2}3} =-\cos \theta .
\end{equation}

We then calculate the superpotential $\sS{}_{\alf} {}^{\mu \nu } = \sS{}_{a} {}^{\mu \nu } h^a {}_{\alpha}$, defined by (\ref{super_K}), for the tetrad \eqref{Lemtet} and the inertial spin connection \eqref{wmlem}
and find the non-vanishing components 
\begin{equation}\label{superLG}
    \begin{array}{cccc}
   \sS{}_{0} {}^{0 1} = - \sS{}_{0} {}^{1 0} = -\frac{4 M}{r^2},\quad \sS{}_{1} {}^{0 1} = - \sS{}_{1} {}^{1 0} = -\frac{2}{r}\sqrt{\frac{2M}{r}}\frac{1}{1-2 M/r},\\

\sS{}_{2} {}^{0 2} = {\sS}{}_{3} {}^{0 3} = - \sS{}_{2} {}^{2 0} = - \sS{}_{3} {}^{3 0} = -\frac{1}{2r}\sqrt{\frac{2M}{r}}\frac{1}{1-2 M/r},\\

\sS{}_{2} {}^{1 2} = {\sS}{}_{3} {}^{1 3} = -\sS{}_{2} {}^{2 1}= - \sS{}_{3} {}^{3 1} = \frac{M}{r^2}.\\

    \end{array}
\end{equation}

Now we have to address the issue of choosing the vector $\xi$ generating the Noether charge. We can consider $\xi$ as a four-velocity of a freely falling observer in the form
\be
\tilde{\xi}^\alpha=(-1,0,0,0),
\m{xi_Lem}
\ee
which corresponds to identification of $\xi$ with the zeroth component of the tetrad, i.e. ${\tilde{\xi}=-\stackrel{C}{h}_{\hat{0}}}$.
While this is formally the same as \eqref{Killingtime}, this vector is in the Lemaitre coordinates and hence it is not a Killing vector.

Using \eqref{xi_Lem}, we obtain a vanishing Noether current
\begin{equation}\label{zerocurr}
{\scJ}{}^\alf(\tilde{\xi}) = \l(0,0,0,0 \r).
\end{equation}

This result can be understood as a consequence of the equivalence principle, i.e. the free-falling observer measures a vanishing Noether energy-momentum current. Consequently  we obtain a zero Noether conserved charge
\begin{equation}\label{zerocharge}
\mathcal{P}(\tilde{\xi})=0,
\end{equation}
 i.e. the free-falling observer measures a vanishing total mass of the black hole, which is the same result as obtained in references \cite{Maluf0704} and \cite{Obukhov+}.

We can also define a new proper frame from \eqref{Lemtet} by applying a local Lorentz transformation  (\ref{GrindEQ__36_}). The inertial spin connection becomes zero and the corresponding proper tetrad takes the form
\begin{equation} \label{Lemproper}
{\stackrel{D}{h}}{}^{a} {}_{\mu } \equiv \left[\begin{array}{cccc} {1} & {0} & {0} & {0} \\ {0} & { \sin \theta \cos \varphi \left(\frac{2M}{r} \right)^{{-1\mathord{\left/ {\vphantom {-1 2}} \right. \kern-\nulldelimiterspace} 2} } } & {r\cos \theta \cos \varphi } & {-r\sin \theta \sin \varphi } \\ {0} & {\sin \theta \sin \varphi \left(\frac{2M}{r} \right)^{{-1\mathord{\left/ {\vphantom {-1 2}} \right. \kern-\nulldelimiterspace} 2} } } & {r\cos \theta \sin \varphi } & {r\sin \theta \cos \varphi } \\ {0} & {\cos \theta \left(\frac{2M}{r} \right)^{{-1\mathord{\left/ {\vphantom {-1 2}} \right. \kern-\nulldelimiterspace} 2} } } & {-r\sin \theta } & {0} \end{array}\right].
\end{equation}
Because our formalism is Lorentz covariant, applying it in the framework of the proper tetrad (\ref{Lemproper}), we obtain again  the vanishing current (\ref{zerocurr}).

Following the logic of the previous section we denote the diagonal tetrad (\ref{Lemtet}) corresponding the Lemaitre metric (\ref{metriclem}) with the inertial spin connection (\ref{wmlem}) as the  {\em Lemaitre gauge}. After any coordinate transformations or local Lorentz transformations applied  simultaneously to (\ref{Lemtet}) and (\ref{wmlem}), the resulting  tetrad and the related inertial spin connection will represent the Lemaitre gauge as well.

 \section{Comparison of Schwarzschild static and Lemaitre gauges}
\setcounter{equation}{0}
\m{5}

The tetrads \eqref{Lemtet} and \eqref{Lemproper} obtained in the Lemaitre gauge are written in the Lemaitre coordinates. In order to make a direct comparison with the  tetrads \eqref{BHtet}  and \eqref{GrindEQ__37_} obtained in the Schwarzschild  static gauge, let us write all  formulae from the Lemaitre gauge in the same  Schwarzschild coordinates. Applying the transformations (\ref{SchtoLemTransf}) to  the tetrad (\ref{Lemtet}), we can write (\ref{Lemtet}) in the  Schwarzschild coordinates as
\begin{equation}\label{LemtetinSch}
{\stackrel{C}{h}}{}^a{}_\mu \equiv
    \left(
\begin{array}{cccc}
 1 & \frac{\sqrt{2Mr}}{r-2M} & 0 & 0 \\
 \sqrt{\frac{2M}{r}} & \frac{r}{r-2M} & 0 & 0 \\
 0 & 0 & r & 0 \\
 0 & 0 & 0 & r \sin\theta \\
\end{array}
\right).
\end{equation}
 The related inertial spin connection (\ref{wmlem}) is not changed because the transformations (\ref{SchtoLemTransf}) affect 0 and 1 coordinates only. Moreover, we can write the Lemaitre proper tetrad (\ref{Lemproper})  in the Schwarzschild coordinates as
 \begin{equation}\label{LemproperinSch}
     {\stackrel{D}{h}}{}^a{}_\mu=\left(
\begin{array}{cccc}
 1 & \frac{\sqrt{2Mr}}{r-2M} & 0 & 0 \\
 \sqrt{\frac{2M}{r}} \cos\varphi \sin\theta & \frac{r \cos\varphi \sin\theta}{r-2M} & r \cos\theta \cos\varphi & -r \sin\theta \sin\varphi \\
 \sqrt{\frac{2M}{r}} \sin\theta \sin\varphi & \frac{r \sin\theta \sin\varphi}{r-2M} & r \cos\theta \sin\varphi & r \cos\varphi \sin\theta \\
 \sqrt{\frac{2M}{r}} \cos\theta & \frac{r \cos\theta}{r-2M} & -r \sin\theta & 0 \\
\end{array}
\right).
 \end{equation}
Of course, the vanishing inertial spin connection related to (\ref{LemproperinSch}) in the Lemaitre gauge is zero as well.

Thus, we have now all four tetrads $\stackrel{A}{h}$  (\ref{GrindEQ__37_}), $\stackrel{B}{h}$ (\ref{BHtet}), $\stackrel{C}{h}$  (\ref{LemtetinSch}) and $\stackrel{D}{h}$ (\ref{LemproperinSch}),  written in the same Schwarzschild coordinates. It turns out that all of these tetrads are connected directly. The tetrad $\stackrel{B}{h}$ in (\ref{BHtet}) can be transformed to the tetrad  $\stackrel{C}{h}$ in (\ref{LemtetinSch})
\be
{\stackrel{C}{h}}{}^a{}_\mu= (\Lambda_{(boost)})^a {}_b{\stackrel{B}{h}}{}^b{}_\mu
\m{B_goto_C},
\ee
  by making a radial boost
\begin{equation}\label{Lamdaboost}
    (\Lambda_{(boost)})^a {}_b =
    \left(
\begin{array}{cccc}
 \gamma & \gamma \beta & 0 & 0 \\
 \gamma \beta & \gamma & 0 & 0 \\
 0 & 0 & 1 & 0 \\
 0 & 0 & 0 & 1 \\
\end{array}
\right).
\end{equation}
where $\gamma=1/ \sqrt{1-\beta^2}$ and $\beta$ is a rapidity given by
\begin{equation}\label{rapidityTEGR}
 \beta= \sqrt{\frac{2M}{r}}.
 \end{equation}
Finally, we conclude that the tetrads in Schwarzschild coordinates named as $\stackrel{A}{h}$, $\stackrel{B}{h}$, $ \stackrel{C}{h}$, $ \stackrel{D}{h}$ are connected by a series of local Lorentz transformations
\begin{equation}\label{alltetradtransform}
   ~~ \stackrel{A}{h} ~~\stackrel{\Lambda_{(Sch)}}{\longrightarrow} ~~\stackrel{B}{h} ~~\stackrel{\Lambda_{(boost)}}{\longrightarrow} ~~\stackrel{C}{h}~~ \stackrel{\Lambda_{(Sch)}^{-1}}{\longrightarrow} ~~\stackrel{D}{h},
\end{equation}
or written explicitly
\begin{equation}\label{att}
    \begin{array}{cccc}
      \stackrel{B}{h}{}\!^a {}_\mu  =(\Lambda_{(Sch)})^a {}_b    \stackrel{A}{h}{}\!^b {}_\mu,\\
      \stackrel{C}{h}\!{}^a {}_\mu =(\Lambda_{(boost)})^a {}_b    \stackrel{B}{h}{}\!^b {}_\mu, \\
    \stackrel{D}{h}{}\!^a {}_\mu  =(\Lambda_{(Sch)}^{-1})^a {}_b    \stackrel{C}{h}{}\!^b {}_\mu. \\
         \end{array}
\end{equation}

Inverse relations are given by the scheme

\begin{equation}\label{alltetradtransform_1}
  ~~ \stackrel{D}{h} ~~\stackrel{\Lambda_{(Sch)}}{\longrightarrow} ~~\stackrel{C}{h} ~~\stackrel{\Lambda^{-1}_{(boost)}}{\longrightarrow} ~~\stackrel{B}{h}~~ \stackrel{\Lambda_{(Sch)}^{-1}}{\longrightarrow} ~~\stackrel{A}{h}.
\end{equation}

Moreover, we can consider a velocity  of the free-falling observer to the black hole. In the Lemaitre coordinates it was $\tilde{\xi}^\alf=(-1,0,0,0)$. In the Schwarzschild coordinates, using (\ref{SchtoLemTransf}), it becomes
\begin{equation} \label{velocityff1}
\tilde{\xi}^\alpha= \l(-\frac{1}{f},\sqrt{1-f},0,0\r) = \l(-\frac{1}{1-2M/r},\sqrt{\frac{2M}{r}},0,0\r).
\end{equation}

\subsection{Different forms of the Schwarzschild static gauge}
\m{5_1}
We could see that all four considered tetrads are related as \eqref{alltetradtransform}. Starting with a tetrad and the corresponding spin connection in the Schwarzschild static gauge, i.e. \eqref{BHtet} and \eqref{BHspin}, and transforming simultaneously both variables we can obtain different representations  of the Schwarzschild static gauge. We can schematically represent this as on the Figure 1, where the dots mean the tetrads and arrows mean the Lorentz rotations applied to both the tetrad and the spin connection, the bold dot means the initial tetrad in which we switch-off gravity

\bigskip
\centerline{\begin{picture}(120,120)
\put(22,40){\vector(1,0){75}}
\put(20,100){\vector(1,0){77}}
\put(20,100){\vector(0,-1){57}}
\put(20,100){\circle*{8}}
\put(20,40){\circle*{3}}
\put(100,40){\circle*{3}}
\put(100,100){\circle*{3}}
\put(5,40){$C$}
\put(5,100){$B$}
\put(105,100){$A$}
\put(105,40){$D$}
\put(105,100){$A({\rm proper})$}
\put(50,105){$\Lambda^{-1}_{(Sch)}$}
\put(22,70){$\Lambda_{(boost)}$}
\put(50,45){$\Lambda^{-1}_{(Sch)}$}
\put(40,10){{\rm Figure 1}}
\end{picture}}
\noindent Thus,  the pairs of tetrads and their related inertial spin connections in the Schwarzschild static gauge are as follows:

 a) The tetrad $\stackrel{A}{h}$ is a proper tetrad with zero inertial spin connection.

b) The tetrad $\stackrel{B}{h}$ and  the inertial spin connection (\ref{BHspin}).

c) The tetrad $\stackrel{C}{h}$ and the inertial spin connection  obtained by a composite local Lorentz transformation consisting of the rotation \eqref{GrindEQ__36_} and  boost \eqref{Lamdaboost}, i.e.
\bea
    \sA{}^a{}_{b\mu}& =&    (\Lambda_{(AC)}){}^a {}_c \partial_\mu    (\Lambda^{-1}_{(AC)}){}^c {}_b, \label{A_C}\\
    (\Lambda_{(AC)}){}^a {}_b &\equiv & ({\Lambda}_{(boost)}){}^a {}_c (\Lambda_{(Sch)}){}^c {}_b.
    \nonumber
\eea
 Non-zero components of (\ref{A_C}) are given in Appendix~\ref{appendA}, see (\ref{h_C_Sch_spin}).

d) The tetrad $\stackrel{D}{h}$ and inertial spin connection  obtained from the composite local Lorentz transformation consisting of the boost (\ref{Lamdaboost}), the rotation (\ref{GrindEQ__36_}) and its inverse, i.e.
\bea
\label{A_D}
    \sA{}^a{}_{b\mu}& =&    (\Lambda_{(AD)}){}^a {}_c \partial_\mu    (\Lambda^{-1}_{(AD)}){}^c {}_b;\\
    (\Lambda_{(AD)}){}^a {}_b &\equiv & (\Lambda^{-1}_{(Sch)}){}^a {}_c (\Lambda_{(boost)}){}^c {}_d (\Lambda_{(Sch)}){}^d {}_b .
    \nonumber
\eea
Non-zero components of (\ref{A_D}) are given in Appendix~\ref{appendA}, see (\ref{ISC_Sch_d}).

For all  pairs of the tetrads $\stackrel{A}{h},\stackrel{B}{h},\stackrel{C}{h},\stackrel{D}{h}$, and their corresponding inertial spin connections in the Schwarzschild static gauge (see Figure 1) the corresponding superpotential is given by the same expression \eqref{superSG} due to Lorentz invariance.
Using the time-like Killing  vector \eqref{Killingtime} in (\ref{Energy}), we find the Noether charge that can be identified as the total black hole mass is given by
\eqref{E=M}, i.e.%
\begin{equation}
    \mathcal{P}(\xi)=M.
\end{equation}

Interestingly, if we consider $\xi$ to be the velocity vector of the free-falling observer  \eqref{velocityff1}  and use it to calculate the total energy in the Schwarzschild static gauge, we still obtain the same result $\mathcal{P}=M$. However, this is just a coincidence on the account of Schwarzschild static gauge being very unique and highly symmetric with \eqref{asymptSchw}. We will discuss this in section~\ref{secdisc}.

\subsection{Different forms of the Lemaitre gauge}
\m{5_2}

In a similar fashion, we can obtain different forms of the Lemaitre gauge. We start with one form of the Lemaitre gauge, i.e. \eqref{LemtetinSch} and \eqref{wmlem}, and apply  transformations in (\ref{alltetradtransform}) simultaneously to both the tetrad and the inertial spin connection. This can be represented schematically by Figure~2, where again the dots mean the tetrads and arrows mean the Lorentz rotations, the bold dot means the initial tetrad in which we switch off gravity:

\bigskip

\centerline{\begin{picture}(120,120)\label{fig2}
\put(22,40){\vector(1,0){75}}
\put(20,100){\vector(1,0){77}}
\put(20,40){\vector(0,1){57}}
\put(20,100){\circle*{3}}
\put(20,40){\circle*{8}}
\put(100,40){\circle*{3}}
\put(100,100){\circle*{3}}
\put(5,40){$C$}
\put(5,100){$B$}
\put(105,100){$A$}
\put(105,40){$D({\rm proper})$}
\put(105,100){$A$}
\put(50,105){$\Lambda^{-1}_{(Sch)}$}
\put(22,70){$\Lambda^{-1}_{(boost)}$}
\put(50,45){$\Lambda^{-1}_{(Sch)}$}
\put(40,10){{\rm Figure 2}}
\end{picture}}
\noindent On the Figure 2, the pairs of tetrads  and their related inertial spin connections in the Lemaitre gauge are as follows:

a) The tetrad $\stackrel{D}{h}$ is a proper one with the vanishing inertial spin connection.

b) The tetrad $\stackrel{C}{h}$  and the inertial spin connection given by (\ref{wmlem}).

c) The tetrad $\stackrel{B}{h}$ and the inertial spin connection calculated from a composite local Lorentz transformation consisting of the rotation \eqref{GrindEQ__36_} and the inverse boost \eqref{Lamdaboost}, i.e.
\bea
    \sA{}^a{}_{b\mu}& =&    (\Lambda_{(DB)}){}^a {}_c \partial_\mu    (\Lambda^{-1}_{(DB)}){}^c {}_b, \label{D_B}\\
    (\Lambda_{(DB)}){}^a {}_b &\equiv & ({\Lambda}^{-1}_{(boost)}){}^a {}_c (\Lambda_{(Sch)}){}^c {}_b.
    \nonumber
\eea
 Non-zero components of (\ref{D_B}) are given in Appendix~\ref{appendA}, see (\ref{h_B_Lem_spin}).

d) The tetrad $\stackrel{A}{h}$ and  the inertial spin connection obtained from a composite local Lorentz transformation consisting of  the rotation (\ref{GrindEQ__36_}), boost (\ref{Lamdaboost}), and the inverse rotation (\ref{GrindEQ__36_}), i.e.
\bea
\label{D_A}
    \sA{}^a{}_{b\mu}& =&    (\Lambda_{(DA)}){}^a {}_c \partial_\mu    (\Lambda^{-1}_{(DA)}){}^c {}_b,\\
    (\Lambda_{(DA)}){}^a {}_b &\equiv & (\Lambda^{-1}_{(Sch)}){}^a {}_c (\Lambda^{-1}_{(boost)}){}^c {}_d (\Lambda_{(Sch)}){}^d {}_b .
    \nonumber
\eea
Non-zero components of (\ref{D_A}) are given in Appendix~\ref{appendA}, see (\ref{h_A_Lem_spin}).

All pairs of the tetrads $\stackrel{D}{h},\stackrel{C}{h},\stackrel{B}{h},\stackrel{A}{h}$ with their related inertial spin connections, as listed a) to d) above, represent the Lemaitre gauge and hence lead to the  same superpotential  given by the expression \eqref{superLG}  due to the Lorentz invariance.

Let us now discuss the choice of the vector $\xi$ and calculation of the Noether charges. If we choose $\xi$ to be given by \eqref{velocityff1} we find  that both the  Noether current \eqref{zerocurr} and conserved charge \eqref{zerocharge} vanish for all combinations of tetrads and inertial spin connections in points a) to d) above.

All of the  tetrads listed above correspond to the same Schwarzschild metric \eqref{BHmet}, despite being in the Lemaitre gauge. Therefore, we can also consider a  time-like Killing vector  of the metric \eqref{BHmet} given by \eqref{Killingtime} and calculate the corresponding Noether charge
\be
\mathcal{P}(\xi)=2M.
\m{E_2M}
\ee
Surprisingly, we obtain a finite value, which is the twice of the result in the Schwarzschild static gauge \eqref{E=M}. We discuss this result in the following section.

\subsection{Discussion \label{secdisc}}

Let us first compare our results with the previous works  by Maluf \textit{et. al.} \cite{Maluf0704} and Lucas \textit{et. al.}
\cite{Obukhov+} in which the same situation with free-falling observers was considered. Both these works have calculated the total energy  $P_{\hat{0}}=0$ rather than the Noether charges. However, as we have discussed at the end of section~\ref{seccharges}, these coincide under certain conditions.  In \cite{Maluf0704}, the authors work in the non-covariant formulation of TEGR where the spin connection is assumed to be identically zero and hence they work only with what we call the proper tetrads. Their tetrad (27) in \cite{Maluf0704} is identical to our tetrad \eqref{LemproperinSch} and they  find  $P_{\hat{0}}=0$ for the zeroth component of the energy-momentum \eqref{Pa} and  identify this correctly as a consequence of the equivalence principle.
In \cite{Obukhov+}, the same situation is considered in the framework of the covariant  formulation and the authors show that the spin connection ``regularizes", i.e. leads to the finite, total energy using arbitrary tetrads. While in \cite{Maluf0704}, the authors claim that in the case of a freely falling frame a possibility to calculate the total mass is lost,  the paper \cite{Obukhov+} shows that it is  possible to obtain $M$ using the free-falling tetrads in the covariant formulation. In our formalism, this is equivalent to using a spin connection corresponding to the tetrad $\stackrel{D}{h}$ in the  Schwarzchild static gauge, i.e. the case d) in section~\ref{5_1}.
On the other hand, references \cite{Obukhov:2006ge, Obukhov_2006,Obukhov_Rubilar_Pereira_2006} calculate the Noether charges, but for the Schwarzschild solution,  only in the Schwarzschild static gauge, leading always to the same answer $M$.

We have calculated the Noether charges for both the Schwarzschild static and Lemaitre gauges and considered various choices for the vector field $\xi$.  Our results  can be summarized  in Table~\ref{table1}:
\begin{table}[h!]
\begin{center}
\begin{tabular}{ |c|c|c|}
 \hline
 Gauge & $\mathcal{P}(\xi)$ & $\mathcal{P}(\tilde{\xi})$  %& $P_{\hat{0}}$
 \\
 \hline
 Schwarzschild static& $\textbf{M}$ & $M$ %& $M$
 \\
 \hline
 Lemaitre & $2M$ & $\textbf{0}$ %& $0$
 \\
 \hline
\end{tabular}
\caption{\small Noether charges in the Schwarzschild static and Lemaitre gauges for different choices of the vector field $\xi$.\label{table1}}
\end{center}
\end{table}

Let us now explain the meaning of each result in the table above. The Schwarzschild static gauge represents the physical configuration of the physical fields corresponding to the static observers. The choice of the vector $\xi$ as \eqref{Killingtime} is a very special one due to the fact that \eqref{Killingtime} is both a Killing vector as well as a velocity of the static observer at infinity  
\begin{equation}
\xi=-\lim_{r\rightarrow \infty}\stackrel{B}{h}_{\hat{0}},
\end{equation}
and hence we  obtain the same $\mathcal{P}(\xi) = M$. Therefore, based on this example, we cannot decide whether $\xi$ should be chosen as a Killing vector or as a velocity of the observer.

On the other hand, the  Lemaitre gauge represent the physical configuration of the physical fields corresponding to the free-falling observers. The choice of the vector $\tilde{\xi}$ as \eqref{velocityff1} represents the velocity of the observer associated with the tetrad \eqref{LemtetinSch}, i.e. $\tilde{\xi}=-\stackrel{C}{h}_{\hat{0}}$, but it is not a Killing vector. In this case, we obtain the physically expected result that the free-falling observer measures the vanishing Noether current and charge. This supports the suggestion to choose $\xi$ to coincide with the observer velocity and hence to associate it with the zeroth component of the tetrad. Therefore, we highlight these two results in the bold face in the table above, in order to stress this fact.

What about the other two results in the table? We can now understand the calculation of $\mathcal{P}(\xi)$ in the Lemaitre gauge as the case when we choose $\xi$ to be a Killing vector of the spacetime metric \eqref{Killingtime} and not a velocity of the observer. As we have shown, this leads to a rather strange result where we obtain the value to be twice of the physical one. This motivates us to conclude that the preferred choice for the vector field $\xi$ is to be a velocity of the observer and not a  Killing vector of the spacetime. However, as we show in the following section, it is possible to recover a correct physical value when we introduce a non-vanishing initial velocity for the observer.

For the sake of completeness, we have also considered $\mathcal{P}(\tilde{\xi})$ in the Schwarzschild static gauge as well. However, here we should be very cautious because $\tilde{\xi}$ is not a Killing vector of the metric \eqref{BHmet}, nor it represent the velocity of the static observer. It is  rather an example of an inconsistent choice of the vector $\xi$ where we attempt to calculate the Noether charge in the static gauge using the velocity of the free-falling observer. While this is obviously not  a meaningful choice, it still leads to the ``correct" answer for the conserved charge. However, this is only a coincidence and produces the correct result only accidentally.

To understand these results from the mathematical viewpoint, let us recall that all conserved quantities are given by the $r\rightarrow\infty$ limit of the superpotential density. In the case of the Schwarzschild static gauge, this has a very special form \eqref{asymptSchw} that explains why we uniquely obtain the result $M$, even in the case when we choose the vector field $\xi$ inconsistently, as we have discussed above. On the other hand,
in the Lemaitre gauge  the leading term expansion of the superpotential \eqref{superLG} is given by
\begin{equation}\label{superLG0}
    \begin{array}{cccc}
   \sS{}_{0} {}^{0 1} = - \sS{}_{0} {}^{1 0} = -\frac{4 M}{r^2},
   \qquad
   \sS{}_{1} {}^{0 1} = - \sS{}_{1} {}^{1 0} = -\sqrt{\frac{2M}{r^3}} +
\mathcal{O}\left(\frac{1}{r^{\frac{5}{2}}}\right)
,\\

\sS{}_{2} {}^{0 2} = {\sS}{}_{3} {}^{0 3} = - \sS{}_{2} {}^{2 0} = - \sS{}_{3} {}^{3 0} = -\sqrt{\frac{M}{2r^3}}  +
\mathcal{O}\left(\frac{1}{r^{\frac{5}{2}}}\right), \\

\sS{}_{2} {}^{1 2} = {\sS}{}_{3} {}^{1 3} = -\sS{}_{2} {}^{2 1}= - \sS{}_{3} {}^{3 1} = \frac{M}{r^2},
    \end{array}
\end{equation}
and hence we have more freedom to contract the superpotential with the components of the vector field $\xi$ and  obtain different values for the Noether charges.

The consequence of the problem above  is the difficulty of interpreting  the total energy-momentum $P_a$ in the covariant formulation. Let us recall that in the Lemaitre gauge both the proper tetrad \eqref{LemproperinSch} and the tetrad \eqref{LemtetinSch} with the corresponding inertial spin connection \eqref{wmlem} should represent the same physical situation,   and hence we would naively expect the same answer for the energy-momentum. This turns out to be not the case. For example, in our case of the Lemaitre gauge, we do obtain the same vanishing zeroth component of $P_{a}$, but if we consider all components of $P_{a}$, we obtain  different results for the other components, i.e.

\begin{equation}
P_a({\stackrel{D}{h}}{}^a{}_\mu,0)=(0,0,0,0), \qquad
P_a({\stackrel{C}{h}}{}^a{}_\mu,\sA^a{}_{b\mu})=(0,\infty,0,0).
\end{equation}

We can understand this result as a consequence of identifying the vector field $\xi$ with the tetrad in the definitions of $P_a$ discussed at the end of section~\ref{seccharges}. While  $\xi$ is  \textit{a priori} arbitrary vector field, in order the Noether charge \eqref{netercharge} to represent a physically meaningful quantity, we need to identify it with the motion of the observer. In the covariant formulation, this poses the problem since the physical configuration of the fields (our gauge) is determined by both the tetrad and the corresponding inertial spin connection, but the motion of the observer is fully given by the tetrad only. Therefore, when we consider different representations of the Lemaitre gauge  discussed in section~\ref{5_2}, each tetrad represents a different observer. We  then encounter the situations where the vector $\xi$ and the tetrad represent two different observers in the same expression for $P_a$, which makes the whole expression unphysical. This is analogous to our unphysical calculation of $\mathcal{P}(\tilde{\xi})$ in the Schwarzschild static gauge discussed above.

Therefore, for the covariant formulation of the theory, it is more meaningful to consider the Noether charges with  the vector field $\xi$ and the tetrad kept independent. In such a setting, the tetrad and inertial spin connection represents the physical configuration of the gravitational field, while the vector field $\xi$ represents an observer  measuring the gravitational energy-momentum. We can argue that physically meaningful results are obtained when both the physical configuration of physical fields and the vector field $\xi$ representing the observer correspond to the same situation. For example, in our case, we choose $\xi$ to be a velocity of the static observer when working in the static gauge, or $\xi$ to be a velocity of the free-falling observer when working in the free-falling Lemaitre gauge.

Note that this is not a problem in the non-covariant formulation where
 the tetrad used in the field equations represents the same observer as the vector field $\xi$ in the Noether conserved charge. These can be identified uniquely with each other and there is no freedom to transform them independently. Therefore, in the non-covariant formulation $P_a$ indeed represents  energy-momentum in a meaningful way.

\section{An arbitrary freely falling observer
\label{secegauge}}
\setcounter{equation}{0}

So far  we have considered  only the case of observers falling to the Schwarzschild black hole from the infinity with the  zero initial velocity. So far, at least to our best knowledge, this was the only case considered in the previous works, e.g. \cite{Maluf0704} and \cite{Obukhov+}. In this section, we generalize this scheme to the case of an arbitrarily free-falling observer, i.e. an observer that start his/her fall to the black hole not from infinity or with a non-vanishing initial velocity.

Solving the geodesic equation in general form for a radially freely falling observer towards the Schwarzschild black hole in the Schwarzschild coordinates (\ref{BHmet}), one obtains
\begin{equation}
\label{velocityff}
\tilde{\tilde{\xi}}^\alpha= \l(-\frac{e}{f},\sqrt{e^2-f},0,0\r)=\l(-\frac{e}{1-2M/r},\sqrt{e^2-(1-2M/r)},0,0\r),
\end{equation}
where $e$ is some number characterizing the initial state of the in-falling particle. The case when $e>1$ corresponds to a nonzero velocity directed to the black hole at infinity $r\goto \infty$, whereas $e<1$ corresponds to a zero velocity at any finite $r_0$. The observer with $e=1$  has a zero velocity at  the infinity $r\goto \infty$,   and the expression \eqref{velocityff} reduces to the (\ref{velocityff1}) considered previously.

\subsection{The case $e >1$}

In this subsection, considering an arbitrary freely falling observer, we  restrict ourselves to the case $e>1$. We construct  the new proper coordinates that are the generalisation of the Lemaitre coordinates (\ref{metriclem}). We then consider a diagonal tetrad and turn off  gravity, like in the previous sections, and find the corresponding inertial spin connection. We then create a new gauge and calculate the corresponding conserved charges.

Keeping in mind (\ref{velocityff}), let us introduce the coordinate transformation
     \bea
          d \tau_e &=& edt +  \frac{\sqrt{e^2-1+\frac{2M}{r}}}{1-\frac{2M}{r}}   dr, \nonumber\\
          d \rho_e &=& dt +   \frac{e}{\left(1-\frac{2M}{r}\right) \sqrt{e^2-1+\frac{2M}{r}}} dr,
          \label{SchToFFcoord}
      \eea
using which, one transforms the Schwarzschild metric (\ref{BHmet}) to the form
 \begin{equation}\label{FFmetric}
     ds^2=-d\tau_e^2 + \left(e^2-1+\frac{2M}{r}\right) d\rho_e^2  + r^2\left( d\theta^2 + \sin^2 \theta d\varphi^2 \right)\,,
 \end{equation}
 where $r = r(\tau_e,\rho_e)$. Using (\ref{SchToFFcoord}), the 4-velocity (\ref{velocityff}) transforms to the form
 \begin{equation}
\tilde{\tilde{\xi}}^\alf = (-1,0,0,0),
     \label{xi_Lem_e}
 \end{equation}
 in the new coordinates \eqref{SchToFFcoord}. Formally it is the  same as (\ref{xi_Lem}) in the coordinates (\ref{metriclem}), but, of  course, in the Schwarzschild coordinates (\ref{metriclem}) the form of (\ref{xi_Lem_e}) changes to \eqref{velocityff}.

The most natural choice for the tetrad corresponding to the metric \eqref{FFmetric} is to take the simplest diagonal tetrad
 \begin{equation}\label{FFdiagtetrad}
      {\stackrel{E}{h}}{}^a{}_\mu=\left(
\begin{array}{cccc}
 1 & 0 & 0 & 0 \\
 0 & \sqrt{e^2+\frac{2M}{r}-1} & 0 & 0 \\
 0 & 0 & r & 0 \\
 0 & 0 & 0 & r \sin (\theta ) \\
\end{array}
\right).
 \end{equation}
To derive the inertial spin connection corresponding to \eqref{FFdiagtetrad} we follow the Lemaitre case discussed in section~\ref{4}, i.e. we calculate the Levi-Civita spin connection
 and ``turn-off"  gravity by $M\goto 0$. We find  the non-vanishing components of the inertial spin connection to be
 \begin{gather}
 \sA{}^{\hat{0}} {}_{\hat{2} 2} =\sA{}^{\hat{2}} {}_{\hat{0} 2} = -\sqrt{e^2-1},
 \quad\sA{}^{\hat{1}} {}_{\hat{2} 2} = - \sA{}^{\hat{2}} {}_{\hat{1} 2} = -e, \quad
 \sA{}^{\hat{0}} {}_{\hat{3} 3} =\sA{}^{\hat{3}} {}_{\hat{0} 3} = -\sin\theta\sqrt{e^2-1},\nonumber
\\
\sA{}^{\hat{1}} {}_{\hat{3} 3} = -\sA{}^{\hat{3}} {}_{\hat{1} 3} = -e \sin\theta,\qquad
\sA{}^{\hat{2}} {}_{\hat{3} 3} = - \sA{}^{\hat{3}} {}_{\hat{2} 3} = -\cos\theta.\label{FFspinc}
 \end{gather}

Analogously to the situation in the Lemaitre case, the tetrad \eqref{FFdiagtetrad} and the inertial spin connection \eqref{FFspinc} are in the new coordinates \eqref{SchToFFcoord}. We can transform them to the Schwarzschild coordinates that allows us more convenient  comparison with the previous results.

We can observe that since the inertial spin connection (\ref{FFspinc}) has 2,3-vector components only, it will have the same form, i.e. \eqref{FFspinc}, in the Schwarzschild coordinates. Applying the coordinate transformation (\ref{SchToFFcoord}) to the diagonal tetrad (\ref{FFdiagtetrad}) we find that  in the Schwarzschild coordinates it will take the form
 \begin{equation}\label{tet_E_Sch}
 {\stackrel{E}{h}}{}^a {}_\mu=   \left(
\begin{array}{cccc}
 e & \frac{\sqrt{e^2+\frac{2M}{r}-1}}{1-\frac{2M}{r}} & 0 & 0 \\
 \sqrt{e^2+\frac{2M}{r}-1} & \frac{e}{1-\frac{2M}{r}} & 0 & 0 \\
 0 & 0 & r & 0 \\
 0 & 0 & 0 & r \sin (\theta ) \\
\end{array}
\right).
 \end{equation}

We can also derive a proper tetrad that we denote  $\stackrel{F}{h}$ by finding a local Lorentz transformation that transforms (\ref{FFspinc}) to zero. We find that this is achieved by
\begin{equation}\label{Lambda_E_F}
    (\Lambda_{(E F)}){}^a {}_b =  (\Lambda_{(Sch)}){}^a {}_c  (\Lambda^{-1}_{(boost')}){}^c {}_b ,
\end{equation}
where
\begin{equation}\label{eboost}
     (\Lambda_{(boost')}){}^a {}_b =
     \left(
\begin{array}{cccc}
 e & -\sqrt{e^2-1} & 0 & 0 \\
 -\sqrt{e^2-1} & e & 0 & 0 \\
 0 & 0 & 1 & 0 \\
 0 & 0 & 0 & 1 \\
\end{array}
\right).
\end{equation}
Applying the Lorentz transformation (\ref{Lambda_E_F}) to \eqref{tet_E_Sch}, $ {\stackrel{F}{h}}{}^a {}_\mu ={\Lambda_{(E F)}}{}^a {}_b  ( \stackrel{E}{h} ){}^b {}_\mu$,  we find the proper tetrad to be
\begin{equation} \label{h_F}
    \begin{array}{cccc}
        {\stackrel{F}{h}}{}^a {}_\mu=
         \left(
\begin{array}{cccc}
 {{\cal A}_e} & \frac{ {{\cal B}_e}}{1-2M/r} & 0 & 0 \\
  \sin\theta  \cos\varphi {{\cal B}_e} & \frac{ \sin\theta \cos\varphi   {{\cal A}_e}}{1-2M/r} & r \cos\theta \cos\varphi
   & -r \sin\theta \sin\varphi \\
 \sin\theta \sin\varphi {{\cal B}_e} & \frac{ \sin\theta \sin\varphi )  {{\cal A}_e}}{1-2M/r} & r \cos\theta \sin\varphi
   & r \sin\theta \cos\varphi \\
 \cos\theta {{\cal B}_e} & \frac{\cos\theta  {{\cal A}_e}}{1-2M/r} & -r \sin\theta & 0 \\
\end{array}
\right),
    \end{array}
\end{equation}
where we have introduced
\bea
{{\cal A}_e} &\equiv & e^2-\sqrt{e^2-1} \sqrt{e^2+\frac{2M}{r}-1},\qquad {{\cal A}_{e=1} = 1},
\m{A_e}\\
{{\cal B}_e} &\equiv & e \left(\sqrt{e^2+\frac{2M}{r}-1}-\sqrt{e^2-1}\right),\qquad {{\cal B}_{e=1} = \sqrt{\frac{2M}{r}}}.
\m{B_e}
\eea

In the spirit of our previous results, we call the tetrad (\ref{tet_E_Sch}) and the corresponding inertial spin connection (\ref{FFspinc}), or alternatively the new proper tetrad \eqref{h_F},  a {\em generalized Lemaitre gauge}, or $e$-gauge.
Moreover, analogously to \eqref{alltetradtransformFF}, we can relate new tetrads $\stackrel{E}{h}$ and $\stackrel{F}{h}$ with the tetrads $\stackrel{B}{h}$ and $\stackrel{A}{h}$ as
\begin{equation}\label{alltetradtransformFF}
   ~~ \stackrel{A}{h} ~~\stackrel{\Lambda_{(Sch)}}{\longrightarrow} ~~\stackrel{B}{h} ~~\stackrel{ \Lambda^{-1}_{(boost_e)}}{\longrightarrow} ~~\stackrel{E}{h}~~ \stackrel{\Lambda_{(EF)}}{\longrightarrow} ~~\stackrel{F}{h},
\end{equation}
where we have introduced the boost
\begin{equation}
 (\Lambda_{(boost_e)}){}^a {}_b =
     \left(
\begin{array}{cccc}
 \frac{e}{\sqrt{1-\frac{2M}{r}}} & \frac{\sqrt{e^2+\frac{2M}{r}-1}}{\sqrt{1-\frac{2M}{r}}} & 0 & 0 \\
 \frac{\sqrt{e^2+\frac{2M}{r}-1}}{\sqrt{1-\frac{2M}{r}}} & \frac{e}{\sqrt{1-\frac{2M}{r}}} & 0 & 0 \\
 0 & 0 & 1 & 0 \\
 0 & 0 & 0 & 1 \\
\end{array}
\right).
 \end{equation}
Taking the tetrad  (\ref{tet_E_Sch}) and the corresponding inertial spin connection (\ref{FFspinc}), and simultaneously transforming both the tetrad and  inertial spin connection using the above transformations, we obtain different forms of the  $e$-gauge. This can be illustrated by the following scheme

\bigskip

\centerline{\begin{picture}(120,120)
\put(22,40){\vector(1,0){75}}
\put(20,100){\vector(1,0){77}}
\put(20,40){\vector(0,1){57}}
\put(20,100){\circle*{3}}
\put(20,40){\circle*{8}}
\put(100,40){\circle*{3}}
\put(100,100){\circle*{3}}
\put(5,40){$E$}
\put(5,100){$B$}
\put(105,100){$A$}
\put(105,40){$F({\rm proper})$}
\put(105,100){$A$}
\put(50,105){$\Lambda^{-1}_{(Sch)}$}
\put(22,70){$\Lambda^{-1}_{(boost_e)}$}
\put(50,45){$\Lambda_{(EF)}$}
\put(40,10){{\rm Figure 5}}
\end{picture}}
\noindent
Or, explicitly we can write that the $e$-gauge is given by following combinations of the tetrad and  inertial spin connection:

a) The tetrad $\stackrel{F}{h}$ is a proper tetrad, i.e. with a zero inertial spin connection.

b) The tetrad $\stackrel{E}{h}$ and the inertial spin connection (\ref{FFspinc}).

c) The tetrad $\stackrel{B}{h}$ and the inertial spin connection obtained by
\bea
\label{A_e_B}
    \sA{}^a{}_{b\mu}& =&    (\Lambda_{(FB)}){}^a {}_c \partial_\mu    (\Lambda^{-1}_{(FB)}){}^c {}_b;\\
    (\Lambda_{(FB)}){}^a {}_b &\equiv & (\Lambda^{-1}_{(boost_e)}){}^a {}_c (\Lambda^{-1}_{(boost')}){}^c {}_d (\Lambda_{(Sch)}){}^d {}_b .
    \nonumber
\eea
The non-zero components of  (\ref{A_e_B}) are listed in Appendix~\ref{appendA}, see (\ref{FFspincinhB}).

d) The tetrad $\stackrel{A}{h}$ and the inertial spin connection
\bea
\label{A_e_A}
    \sA{}^a{}_{b\mu} &=&    (\Lambda_{(FA)}){}^a {}_c \partial_\mu    (\Lambda^{-1}_{(FA)}){}^c {}_b;\\
    (\Lambda_{(FA)}){}^a {}_b &\equiv & (\Lambda^{-1}_{(Sch)}){}^a {}_e (\Lambda^{-1}_{(boost_e)}){}^e {}_c (\Lambda^{-1}_{(boost')}){}^c {}_d (\Lambda_{(Sch)}){}^d {}_b .
    \nonumber
\eea
The non-zero components of  (\ref{A_e_A}) are listed in Appendix~\ref{appendA}, see (\ref{FFspincinhA}).

Due to the Lorentz invariance,  all the pairs $\stackrel{F}{h},\stackrel{E}{h},\stackrel{B}{h},\stackrel{A}{h}$ with their related inertial spin connections  listed above as a)-d) lead to the same result for the superpotential
\begin{equation}\label{FFsuper}
    \begin{array}{cccc}
  \sS{}_{{0}} {}^{{0} 1} = -\sS{}_{{0}} {}^{{1} 0} = -\frac{2}{r}  \left({\cal A}_e -1 +2M/r\right),
\quad
\sS{}_{{1}} {}^{{0} 1} = - \sS{}_{{1}} {}^{{1} 0} = \frac{2}{r} \frac{{\cal B}_e}{1 - 2M/r},
\\
\sS{}_{{2}} {}^{{0} 2} = \sS{}_{{3}} {}^{{0} 3} = - \sS{}_{{2}} {}^{{2} 0} = - \sS{}_{{3}} {}^{{3} 0} =  \frac{e}{r}\frac{{\cal A}_e -1 +M/r}{(1 - 2M/r)\sqrt{e^2 +2M/r -1})},\\

\sS{}_{{2}} {}^{{1} 2} = \sS{}_{{3}} {}^{{1} 3} = - \sS{}_{{2}} {}^{{2} 1} = - \sS{}_{{3}} {}^{{3} 1} = -\frac{1}{r}  \left({\cal A}_e -1 +M/r\right),\\
    \end{array}
\end{equation}
where ${\cal A}_e$ and ${\cal B}_e$ are defined in (\ref{A_e}) and (\ref{B_e}), respectively.

Let us now calculate the Noether current (\ref{FFsuper}) and the corresponding Noether charges \eqref{netercharge} for different vector fields $\xi$. For the vector field $\tilde{\tilde{\xi}}^\alf$ chosen to represent the arbitrary freely-falling observer (\ref{velocityff}) we find the vanishing Noether current
\bea
    {\scJ}{}^\alf(\tilde{\tilde{\xi}}) &=&\l( 0, 0, 0, 0\r)\,
  \label{currbh_1}
\eea
and hence naturally leads to the vanishing Noether conserved charge
\begin{equation}
\mathcal{P}(\tilde{\tilde{\xi}})=0,
\label{currbh_E}
\end{equation}
what is analogous to the situation in the Lemaitre case and is consistent with the expectations based on the equivalence principle.
However, an interesting new result is that if  we choose the vector $\xi$ as the time-like Killing vector of the metric \eqref{BHmet} given by \eqref{Killingtime}, we obtain
\begin{equation}\label{M_e}
    \mathcal{P}({\xi})=M.
\end{equation}
\newpage We can summarize our results in the following table\footnote{If we consider the vector field $\xi$ to be the Lemaitre free-falling vector (\ref{velocityff1}), we obtain naively a good result $\mathcal{P}(\tilde{\xi})=M$ as well. However, this suffers from the same problem as the previous calculation of $\mathcal{P}(\tilde{\xi})=M$ in the static gauge discussed in section~\ref{secdisc}, i.e. the vector $\tilde{\xi}$ is  neither a  Killing vector nor a velocity of the observer in the $e$-gauge, what makes the physical interpretation of this result rather unclear. Therefore, we consider this result to be unphysical. We highlight the physical results in the bold face in the Table~\ref{tab2}. }:
\begin{table}[h!]
\begin{center}
\begin{tabular}{ |c|c|c|c|}
 \hline
 Gauge & $\mathcal{P}(\xi)$ & $\mathcal{P}(\tilde{\xi})$  & $\mathcal{P}(\tilde{\tilde{\xi}})$  %& $P_{\hat{0}}$
 \\
 \hline
 $e$-gauge& $\textbf{M}$ & M & $\textbf{0}$ %& $M$
 \\
 \hline
\end{tabular}
\caption{{\small Noether charges in the e-gauge for different choices of the vector field $\xi$.}}\label{tab2}
\end{center}
\end{table}

This is indeed a surprising result since, unlike the Lemaitre case where we found \eqref{E_2M}, we have now the same physical outcome as in the Schwarzschild static gauge.  This means that introducing an initial velocity for the observer changes the result from the curious $2M$ to the physically meaningful value $M$.  We will discuss this result in the following section.

\subsection{Discussion \label{secdiscegauge}}
The generalized Lemaitre gauge or $e$-gauge is the new result of this paper that was not considered before. In the case when we choose the vector field $\xi$ to be associated with the velocity of the observer, i.e. $\xi=-\stackrel{E}{h}_{\hat{0}}$, we  find the vanishing Noether current and charge. This is analogous to the ordinary Lemaitre gauge and can be understood as a consequence of the equivalence principle.

However, an interesting new result is that if we choose the vector $\xi$ to be the time-like Killing vector \eqref{Killingtime}, we find the same result as in the case of the Schwarzschild static gauge.
This means that introducing an initial velocity for the observer  regularizes the corresponding conserved charge from \eqref{E_2M} to \eqref{M_e}. Even more intriguing is that  the change from \eqref{E_2M} to \eqref{M_e} is discontinuous and occurs for an arbitrary small $e$.

While from the physical perspective it is rather unclear how an inclusion of an arbitrary small velocity can regularize the total energy, on the purely mathematical level this can be  understood as a consequence of the regularizing property of the initial velocity $e$. Here we mean the fact that for the standard Lemaitre gauge, the metric \eqref{metriclem} and the corresponding tetrad \eqref{Lemtet} are singular in the limit $M\rightarrow 0$ (or $r\rightarrow\infty$) that is used to determine the inertial spin connection \eqref{wmlem}. On the other hand, the metric \eqref{FFmetric} and the corresponding tetrad \eqref{FFdiagtetrad} are regular in this limit.

The result above is even more curious if we notice an important role of the constant boost \eqref{eboost}. Let us consider the chain of transformations  \eqref{alltetradtransformFF} from  $\stackrel{A}{h}$, which is the proper tetrad in the Schwarzschild static gauge, to $\stackrel{F}{h}$, which is the proper tetrad in the $e$-gauge. The constant boost \eqref{eboost} is contained in the last transformation in the term $\Lambda_{(E F)}$ given by \eqref{Lambda_E_F}, and, as it turns out, is crucial for obtaining the conserved charges discussed above.

The reason why we find this curious is that the constant boost \eqref{eboost} is a global Lorentz transformation that we would not expect to influence any relevant scalar conserved quantities. We remark that  this is not a question of using the covariant or non-covariant formulations, because the whole question of covariance traditionally concerns local Lorentz transformations only, not the global ones.

For the sake of completeness, let us  briefly discuss the case $e<1$ when freely falling observers have ``orbits'' constrained by the finite radius $r_0={2M}/({1-e^2})$. The problem here is that our formula for the Noether charges \eqref{netercharge} relies on taking the asymptotic limit and hence it is not possible to straightforwardly apply it to this case where we have the maximum radius. It is formaly possible to do an analytic continuation to $r>r_0$, leading to  complex quantities $\sT{}^\alpha {}_{\mu\nu}$ and $\sS{}^\alpha {}_{\mu\nu}$. It turns out that the  $\sS{}_{{0}} {}^{{0} 1}$ component of the superpotential   is real at $r>r_0$ and hence we obtain the real Noether charge $\mathcal{P}=M$.  However, despite the fact that one obtains formally acceptable real result in this particular case, the occurrence of complex parameters in intermediate calculations makes the physical interpretation of such results rather unclear \cite{EPT19}. Note that complex quantities in tetrad calculus do appear in the literature, e.g., in \cite{Hohmann:2019nat,EPT19} in the case of cosmological models.

\section{Free-falling tetrads in $f(T)$ gravity\label{secfT}}
\setcounter{equation}{0}
While the primary objective of this paper was to understand conserved charges in TEGR, it is interesting to make analogous calculations in the framework of $f(T)$ gravity. The  $f(T)$ gravity model is one of the most popular modified gravity models where we take the Lagrangian to be an arbitrary function of the torsion scalar \eqref{torscalar} from the TEGR Lagrangian \eqref{lag} \cite{Ferraro:2006jd,Bengochea:2008gz,Ferraro:2008ey,Linder:2010py}
\begin{equation}
\sL_f=\frac{h}{2\kappa} f(\sT).
\end{equation}

Much of the attention was dedicated to finding the solutions of the antisymmetric part of the field equations. Those tetrads that solve the antisymmetric field equations with a vanishing spin connection are known as \textit{good tetrads} \cite{Tamanini:2012hg} or alternatively we can always calculate the corresponding spin connection to arbitrary tetrads in the covariant formulation \cite{Krssak:2015oua,REV_2018}.

To understand the importance of the antisymmetric part of the field equations, let us consider the fully Lorentz-indexed version of the field equations \eqref{FE} in the $f(T)$ case, and write the corresponding Euler-Lagrange expressions $E_{ab}=\eta_{bc}h^c{}_\rho E_a{}^\rho$ as \cite{REV_2018}
\begin{equation}\label{ftequationG}
E_{ab}=h\left(f_{TT} \sS_{ab}{}^{\nu} \partial_{\nu} \sT+f_T \cGG_{ab} +\frac{1}{2} \eta_{ab}(f-\sT\, f_T)\right),
\end{equation}
where $\cGG_{ab}$ is the Levi-Civita Einstein tensor. From this form of the field equations it is clear that the first term is  crucial for understanding the difference of the dynamics of the theory compared to GR. It is also the only term that contributes to the antisymmetric part of the field equations
\begin{equation}\label{fteqas}
E_{[ab]}=f_{TT}\sS_{[ab]}{}^\nu \partial_\nu \sT =0.
\end{equation}
Therefore, the antisymmetric field equation \eqref{fteqas} contain the essential information about the genuine $f(T)$ dynamics what motivates our interest in them. Moreover, it can be shown in the covariant formulation that  the variation with respect to spin connection lead exactly to this antisymmetric part of the tetrad field equations and hence contains the same information \cite{Golovnev:2017dox}.

We can observe that $f_{TT}=0$ trivially satisfies these equations, but this is on the account of reducing the theory to the TEGR case. In a similar fashion, the case $\sT=\text{const}$ reduces the theory to GR with an effective cosmological constant \cite{Ferraro:2011ks,Bejarano:2014bca}. We are primarily interested in the non-trivial solutions of these antisymmetric equations since these represent the potentially genuinely new solutions in $f(T)$ gravity.

The general spherically symmetric spacetime is given by the metric ansatz
\begin{equation}\label{fTmet}
g_{\mu\nu}=\text{diag}(-A^2,B^2,r^2, r^2\sin^2\theta),
\end{equation}
where $A,B$ are some functions of $r$ that in the case of TEGR reduce to \eqref{BHmet}.

We can define various tetrads analogously to TEGR tetrads discussed previously where we keep the same notation but we add a bar above the quantity to distinguish it from the TEGR case, i.e. the diagonal tetrad we denote as
\begin{equation} \label{fTtetdiag}
\stackrel{\bar{B}}{h}{}^{a} {}_{\mu } \equiv {\diag}{\left(A, ~B,~ r, ~r\sin \theta \right)}\, .
\end{equation}
While in the case of TEGR such tetrads lead to divergent conserved charges, in the $f(T)$ case we face the problem of non-vanishing antisymmetric field equations that can be satisfied only when $f_{TT}=0$ and hence reduce trivially to TEGR, i.e. \eqref{fTtetdiag} being a \textit{bad tetrad} \cite{Tamanini:2012hg}.

The solution to this problem is that we have to either consider a \textit{good tetrad} \cite{Tamanini:2012hg}
\begin{equation}\label{tetfTprop}
\stackrel{\bar{A}}{h}{}^{a} {}_{\mu }=(\Lambda_{(Sch)}^{-1}) ^{a} {}_{b} \stackrel{\bar{B}}{h}{}^{b} {}_{\mu }
\end{equation}
where $(\Lambda_{(Sch)}^{-1}) ^{a} {}_{b}$ is given by \eqref{GrindEQ__36_} or  use the  tetrad \eqref{fTtetdiag} with the corresponding spin connection, which turns out to be  given by \eqref{BHspin} \cite{Krssak:2015oua,REV_2018}.

We can observe that in this  particular case  the solution of the problem of the antisymmetric field equations in $f(T)$ gravity seem to be equivalent to the problem of finding conserved charges in TEGR\footnote{Another interesting case where this happens is the case of spatially-flat FRWL spacetime where the diagonal Cartesian tetrad is both the proper and good tetrad}. It is therefore interesting to consider analogues of  accelerated tetrads from the previous sections to demonstrate the differences between these two concepts.

In particular, let us focus on a free-falling tetrad for a general spherically symmetric spacetime \eqref{fTmet}. In the TEGR case, we could consider a radial boost \eqref{Lamdaboost} with the rapidity given by \eqref{rapidityTEGR}, but this requires knowledge of the solution of the field equations $A=1/B=f^{\frac{1}{2}}$, which is \textit{a priori} unknown.

Therefore, in the $f(T)$ case, we must consider a free-falling tetrad for the general spherically symmetric spacetime \eqref{fTmet}. To find it, let us  define the generalized acceleration object $\phi^a{}_b=\cA{}^{\hat{a}}{}_{\hat{b}\hat{0}}$ \cite{Obukhov+}. For the diagonal tetrad \eqref{fTtetdiag}, we find that the only non-vanishing component is $\phi^0{}_1=A'/(AB)$ and hence this diagonal tetrad represents a static observer. We can then consider a  boost in $r$-direction \eqref{Lamdaboost}  with a general rapidity $\bar{\beta}$ and find the non-vanishing generalized acceleration object is transformed as
\begin{equation}
\phi^0{}_1=\phi^1{}_0=\frac{\bar{\gamma}^2 A' +A\bar{\beta}\bar{\beta}'}{AB\bar{\gamma}^3},
\end{equation}
where the prime denotes the derivative with respect to the radial coordinate.

Setting this to zero we find that the generalized acceleration vanishes for the rapidity  $\beta=\pm\sqrt{1-c_1 A^2}$. Without  loss of generality we can choose the solution with a plus sign and choose $c_1=1$, i.e.
\begin{equation}\label{rapfT}
\bar{\beta}=\sqrt{1-A^2}.
\end{equation}

We can  define a  free-falling tetrad for the spherically symmetric spacetime \eqref{fTmet} by
\be
\stackrel{\bar{C}}{h}{}^a{}_\mu= \bar{\Lambda}_{(boost)} {}^a {}_b\stackrel{\bar{B}}{h}{}^b{}_\mu,
%\m{B_goto_C}
\ee
where the $\bar{\Lambda}_{(boost)} {}^a {}_b$ is the radial boost \eqref{Lamdaboost} with the rapidity \eqref{rapfT}.

Let us then attempt to define a $f(T)$ analogue of the proper Lemaitre tetrad \eqref{Lemproper} by combining this radial boost and the proper local Lorentz transformation \eqref{GrindEQ__36_}, i.e
\begin{equation}\label{tetfTD}
\stackrel{\bar{D}}{h}{}^a{}_\mu= (\Lambda_{(Sch)}^{-1}) ^{a} {}_{b} \stackrel{\bar{C}}{h}{}^b{}_\mu = (\Lambda_{(Sch)}^{-1}) ^{a} {}_{b}\bar{\Lambda}_{(boost)} {}^b {}_c\stackrel{\bar{B}}{h}{}^c{}_\mu.
\end{equation}

We can then proceed to calculate the field equations for $\eqref{tetfTD}$ and the zero inertial spin connection  and we find that their antisymmetric part \eqref{fteqas} is given by  
\begin{eqnarray}
E_{[\hat{0}\hat{1}]}&=& -\frac{2 f_{TT} \bar{\beta}\sT'}{rAB} \cos\varphi\sin\theta.\label{ftast0}\\
E_{[\hat{0}\hat{2}]}&=& \frac{2 f_{TT} \bar{\beta}\sT'}{rAB} \cos\theta,\label{ftast1}\\
E_{[\hat{0}\hat{3}]}&=&- \frac{2 f_{TT} \bar{\beta}\sT'}{rAB} \sin\varphi\sin\theta\label{ftast3},
\end{eqnarray}
from where we can immediately identify the problem and see why this construction does not work in the case of $f(T)$ gravity.  These equations are satisfied only if $f_{TT}=0$, $\bar{\beta}=0$ implying $A=1$, or $\sT'=0$, each case reducing the theory to its TEGR limit\footnote{In principle, we could also consider the case $A=1$ with arbitrary $B$. However, as we show in Appendix~\ref{apptriv}, the symmetric field equations imply then $B=1$ as well. Therefore, this case is indeed trivial and does not produce any new sensible good tetrad for $f(T)$ gravity.}

To understand the origins of this result, let us consider the  torsion scalar for both the TEGR solution \eqref{Lemproper} and  the generalized analogue \eqref{tetfTD}. While in the former case we have
\begin{equation}\label{Tzer}
\sT(\stackrel{D}{h}{}^{a}_{\ \mu},0)=0,
\end{equation}
in the latter case we find that
\begin{equation}
\sT(\stackrel{\tilde{D}}{h}{}^{a}_{\ \mu},0)=-
2\frac{A(1+B^2)-2B+2rA'}{r^2AB^2},
\end{equation}
which is generally non-zero and reduces to \eqref{Tzer} only in the case of the TEGR solution $A=1/B=f^{\frac{1}{2}}$.

From here we can see that the radial boost \eqref{Lamdaboost} with \eqref{rapidityTEGR} in the TEGR case has a special role as it makes the tetrad both free-falling and corresponding torsion scalar vanishing. On the other hand, in the case of a general tetrad \eqref{fTtetdiag} the radial boost with the rapidity \eqref{rapfT} makes only the observer free-falling but does not make the corresponding torsion scalar vanishing, what makes the tetrad \eqref{tetfTD} a bad tetrad at the end.

Let us now explain why this deserves our attention.  In the TEGR case, the free-falling tetrad \eqref{Lemproper} gives us a finite conserved charge, either \eqref{zerocharge} or  \eqref{E_2M}, and hence can be considered a proper tetrad. However, as we have just demonstrated, an analogous construction of a generalized free-falling tetrad \eqref{tetfTD} is not a good tetrad in $f(T)$ gravity. Therefore, this is an interesting illustration of the fact that good tetrads in $f(T)$ gravity are different from proper tetrads in TEGR, and they cannot be always constructed as  simple generalizations of the TEGR situation.

For the sake of completeness of our analysis, we can consider a boost that makes the torsion scalar to vanish by generalizing the approach introduced in \cite{Ferraro:2011ks,Bejarano:2014bca}. We consider a time-dependent radial boost with rapidity
\begin{equation}
\beta^*=-\tanh\left(
c_1-\frac{t(A+2rA')}{2rB}
\right)
\end{equation}
and define
\begin{equation}\label{tetGTD}
\stackrel{\bar{G}}{h}{}^a{}_\mu= \bar{\Lambda}_{(boost)} {}^a{}_{b}(\beta^*)  \stackrel{\bar{B}}{h}{}^b{}_\mu
\end{equation}
for which the torsion scalar is vanishing
\begin{equation}
\sT(\stackrel{\bar{G}}{h}{}^{a}_{\ \mu},0)=0.
\end{equation}

While we have not obtained here any new good tetrad, we have demonstrated how to find the most general vanishing torsion scalar solution for the general spherically symmetric spacetime \eqref{fTmet}. In this sense, our result \eqref{tetGTD} is an interesting  new result in the line of previous works
\cite{Ferraro:2011ks,Bejarano:2014bca,Bejarano:2017akj}.

\section{Concluding remarks \label{secdiscfin}}
\setcounter{equation}{0}

In the present paper we have considered several questions related to the the problem of local Lorentz degrees of freedom and the concept of proper and good tetrads  in TEGR and $f(T)$ gravity. The main results concerning various definitions of conserved charges for the static and variously free-falling observers were discussed in length in sections~\ref{secdisc} and \ref{secdiscegauge}, where we have demonstrated how to obtain some physical results and compared our calculations with the previous results  \cite{Maluf0704,Obukhov+}. In particular, we would like to call attention to our discussion of the relation between Noether charges and other energy-momentum definitions used previously, and  generalization of the Lemaitre observers to the case with arbitrary initial velocity named \textit{e-gauge} introduced in Section~\ref{secegauge}.

In these concluding remarks, let us focus on other consequences of our calculations. We recall that there are two formulations of teleparallel theories: non-covariant and covariant. In the non-covariant formulation of TEGR, the field equations are locally Lorentz invariant but various quantities such as the conserved charges and energy-momentum  are not. There is then a class of ``preferred" tetrads called proper tetrads that lead to finite conserved charges. In the covariant formulation, on the other hand, considering an inertial spin connection, we are allowed to use an arbitrarily Lorentz transformed tetrad and make the theory covariant. However, we need to calculate the corresponding spin connection and hence essentially we need to determine the same local Lorentz degrees of freedom as when determining the proper tetrad.

In $f(T)$ gravity, these local Lorentz degrees of freedom play even more more important role since they  affect  the field equations.  In the original non-covariant formulation of $f(T)$ gravity, the field equations are consistent only for the peculiar class of tetrads  known  as {\em good} tetrads \cite{Tamanini:2012hg}. For the Schwarzschild and FRWL geometries  we  find that good tetrads  coincides with  proper tetrad from TEGR. In the covariant formulation, we consider a non-vanishing spin connection that allows us to use arbitrarily Lorentz rotated tetrad in the field equations  and hence we can  ``restore''  the local Lorentz symmetry. However, this local Lorentz symmetry is restored only after we determine the correct spin connection corresponding to the tetrad. Therefore, same as in the case of TEGR, there is need to determine the Lorentz degrees of freedom in both formulations.

There are various approaches to determining these preferred tetrads, or alternatively the corresponding spin connections \cite{Tamanini:2012hg,Toporensky:2019mlg,Hohmann:2019nat,Jarv:2019ctf}. We have considered here the approach of switching off gravity introduced in \cite{Obukhov+} and further developed in \cite{Krssak:2015rqa,Krssak:2015lba,Krssak:2015oua,EPT19,EPT_2020}. While this was known to work very well when  starting with the standard static form of the spherically symmetric metric, we have considered the situation where the starting point is the metric in the Lemaitre coordinate system. Note that this is reasonable since the coordinate system is not determined and hence we should be able to obtain meaningful results using arbitrary coordinate systems as a starting point.

An interesting observation from our calculations is the coincidence between the spin connections calculated starting from the Schwarzschild static coordinates (4.5) and the spin connection calculated  in the Lemaitre coordinates (5.5). Both spin connections were constructed in the same way by starting with the metric, taking the simplest diagonal tetrad, and switching-off gravity. The difference was that the starting points were the metric in different coordinate systems. So, despite the obvious fact that in order to transform from the diagonal Schwarzschild static tetrad to the diagonal Lemaitre tetrad it is necessary to apply a Lorentz boost, the spin connections remain unchanged. The fact that both spin connections (4.5) and (5.5) are the same demonstrates that the process of determination of the spin connection is not unique.

We can understand the origin of this situation in two ways. First it can be viewed as a deficiency of the switching off gravity method used to determine the spin connection. Indeed this method is completely insensitive to any local Lorentz transformation that is proportional to the parameter that controls the strength of gravity since it gets ``switched off". In our case, the boost \eqref{rapidityTEGR} is proportional to $M$ and then  we take the limit $M\goto 0$ to determine the inertial spin connection. Therefore, it is  not a surprise that the resulting spin connection is insensitive to this boost. 

However, we would like to argue that this could be viewed as  a simple demonstration of the problem of  the so-called \textit{remnant symmetries} discovered in $f(T)$ gravity. These were first encountered  in the FRWL spacetime in $f(T)$ gravity, where it was noticed that there exists a class of local Lorentz symmetries that transform a good tetrad into another good tetrad \cite{Ferraro:2014owa,Chen:2014qtl}. This means that despite a manifest violation of local Lorentz invariance in the non-covariant formulation, not all symmetries are violated and some of them remain. This means that good tetrads, and analogously proper tetrads in TEGR, are not unique and are indeed an equivalence class of tetrads related by remnant symmetries. In the covariant formulation, this means that there can exist two tetrads related by a remnant symmetry, and to both of them corresponds the same spin connection, which is the situation we have encountered here. See the recent discussion of Kerr solution \cite{Maluf:2018coz}, where it was demonstrated that two Kerr tetrads share the same connection. For the use of these symmetries in analysis of the problem of degrees of freedom in $f(T)$ gravity see \cite{Golovnev:2020nln}\footnote{For the detailed discussion see our upcoming paper \cite{Krssaktbp}}.

From this follows that we need to pay closer attention to the problem of local Lorentz degrees of freedom in TEGR and the search for a more precise definition of a proper tetrad should be continued. We should also try to understand the relation between the concepts of a proper tetrad in TEGR and a good tetrad in  $f(T)$ theory. We have demonstrated that when we attempt to construct a good tetrad in analogy with the free-falling proper tetrad, we fail, indicating  that proper and good tetrads are in fact different.

\bigskip

{\bf Acknowledgments.} AP has been supported by the Interdisciplinary Scientific and Educational School of Moscow University ``Fundamental and Applied Space Research''; EE and AT are supported by RSF grant 21-12-00130; MK is supported  by  the  CUniverse  research  promotion  initiative (CUAASC) of the Chulalongkorn University.

\appendix

\newpage
\section{Inertial spin connections in various gauges\label{appendA}}
\setcounter{equation}{0}

In this Appendix we derive components of inertial spin connections in various gauges considered in the paper which are  rather lengthy and  cumbersome.

\subsection{The Schwarzschild static gauge}

1) The inertial spin connection corresponding to $\stackrel{C}{h}$ has non-zero components
\begin{gather}
\sA{}^{\hat{0}} {}_{\hat{1} 1} = \sA{}^{\hat{1}} {}_{\hat{0} 1} = \frac{1}{2rf}\sqrt{\frac{2M}{r}},
\quad
\sA{}^{\hat{0}} {}_{\hat{3} 3} = \sin\theta \sA{}^{\hat{0}} {}_{\hat{2} 2} = \sA{}^{\hat{3}} {}_{\hat{0} 3} = \sin\theta \sA{}^{\hat{2}} {}_{\hat{0} 2} = - \frac{1}{f}\sqrt{\frac{2M}{r}}\sin\theta,
\nonumber\\
\sA{}^{\hat{1}} {}_{\hat{3} 3} = \sin \theta \sA{}^{\hat{1}} {}_{\hat{2} 2} = \sA{}^{\hat{3}} {}_{\hat{1} 3} = \sin \theta  \sA{}^{\hat{2}} {}_{\hat{1} 2} = -  \frac{\sin\theta}{f^{\frac{1}{2}}},
\quad 
\sA{}^{\hat{2}} {}_{\hat{3} 3} = -\sA{}^{\hat{3}} {}_{\hat{2} 3} = -\cos\theta. \label{h_C_Sch_spin}
\end{gather}

2) To the tetrad $\stackrel{D}{h}$ corresponds an inertial spin connection with the following non-zero components:

\begin{gather}
   \sA{}^{\hat{0}} {}_{\hat{1} 1} = \sA{}^{\hat{1}} {}_{\hat{0} 1} =
   -\sA{}^{\hat{0}} {}_{\hat{2} 3} = -\sA{}^{\hat{2}} {}_{\hat{0} 3} = \frac{1}{2rf} \sqrt{\frac{2M}{r}} \sin\theta\cos\varphi,
\nonumber \quad
\sA{}^{\hat{0}} {}_{\hat{1} 2} = \sA{}^{\hat{1}} {}_{\hat{0} 2} = - 
\frac{1}{2rf} \sqrt{\frac{2M}{r}} \cos\theta\cos\varphi,
\\
\sA{}^{\hat{0}} {}_{\hat{1} 3} = \sA{}^{\hat{1}} {}_{\hat{0} 3} =\sA{}^{\hat{0}} {}_{\hat{2} 2} = \sA{}^{\hat{2}} {}_{\hat{0} 2}= \frac{1}{2rf} \sqrt{\frac{2M}{r}} \sin\theta\sin\varphi,
\quad
\sA{}^{\hat{0}} {}_{\hat{2} 2} = \sA{}^{\hat{2}} {}_{\hat{0} 2} = -\frac{1}{2rf} \sqrt{\frac{2M}{r}} \cos\theta\sin\varphi,
\nonumber\\
\sA{}^{\hat{0}} {}_{\hat{3} 1} = \sA{}^{\hat{3}} {}_{\hat{0} 1} = \frac{1}{2rf} \sqrt{\frac{2M}{r}} \cos\theta,
\quad
\sA{}^{\hat{0}} {}_{\hat{3} 2} = \sA{}^{\hat{3}} {}_{\hat{0} 2} = 
\frac{1}{2rf} \sqrt{\frac{2M}{r}} \sin\theta,
\nonumber
\\
\sA{}^{\hat{1}} {}_{\hat{2} 3} = - \sA{}^{\hat{2}} {}_{\hat{1} 3} = \left(1- \frac{1}{f}\right)\sin^2\theta,
\quad
\sA{}^{\hat{1}} {}_{\hat{3} 2} =- \sA{}^{\hat{3}} {}_{\hat{1} 2} = - \left(1- \frac{1}{f}\right)\cos\varphi,
\label{ISC_Sch_d}\\
\sA{}^{\hat{1}} {}_{\hat{3} 3} = - \sA{}^{\hat{3}} {}_{\hat{1} 3} =
\left(1- \frac{1}{f}\right)\sin\theta\cos\theta\sin\varphi,
\quad
\sA{}^{\hat{2}} {}_{\hat{3} 2} = - \sA{}^{\hat{3}} {}_{\hat{2} 2} = -
\left(1- \frac{1}{f}\right)\sin\varphi,
\nonumber\\
\sA{}^{\hat{2}} {}_{\hat{3} 3} = - \sA{}^{\hat{3}} {}_{\hat{2} 3} =  -\left(1- \frac{1}{f}\right)\sin\theta\cos\theta\cos\varphi. \nonumber
\end{gather}

\subsection{The Lemaitre gauge}

1) To the tetrad $\stackrel{B}{h}$ corresponds an inertial spin connection with the following non-zero components:
\begin{gather}
\sA{}^{\hat{0}} {}_{\hat{1} 1} = \sA{}^{\hat{1}} {}_{\hat{0} 1} =- \frac{1}{2rf}\sqrt{\frac{2M}{r}},
\quad
\sA{}^{\hat{0}} {}_{\hat{3} 3} = \sin\theta \sA{}^{\hat{0}} {}_{\hat{2} 2} = \sA{}^{\hat{3}} {}_{\hat{0} 3} = \sin\theta \sA{}^{\hat{2}} {}_{\hat{0} 2} = \frac{1}{f}\sqrt{\frac{2M}{r}}\sin\theta,
\nonumber\\
\sA{}^{\hat{1}} {}_{\hat{3} 3} = \sin \theta \sA{}^{\hat{1}} {}_{\hat{2} 2} = \sA{}^{\hat{3}} {}_{\hat{1} 3} = \sin \theta  \sA{}^{\hat{2}} {}_{\hat{1} 2} = -  \frac{\sin\theta}{f^{\frac{1}{2}}},
\quad 
\sA{}^{\hat{2}} {}_{\hat{3} 3} = -\sA{}^{\hat{3}} {}_{\hat{2} 3} = -\cos\theta,  \label{h_B_Lem_spin}
\end{gather}
which we can recognize to be almost identical to \eqref{h_C_Sch_spin}, but the components with $\hat{0}$ have a minus sign.

2) To the tetrad $\stackrel{A }{h}$ corresponds an inertial spin connection with the following non-zero components:
\begin{gather}
   \sA{}^{\hat{0}} {}_{\hat{1} 1} = \sA{}^{\hat{1}} {}_{\hat{0} 1} =
   -\sA{}^{\hat{0}} {}_{\hat{2} 3} = -\sA{}^{\hat{2}} {}_{\hat{0} 3} = -\frac{1}{2rf} \sqrt{\frac{2M}{r}} \sin\theta\cos\varphi,
\nonumber \quad
\sA{}^{\hat{0}} {}_{\hat{1} 2} = \sA{}^{\hat{1}} {}_{\hat{0} 2} = 
\frac{1}{2rf} \sqrt{\frac{2M}{r}} \cos\theta\cos\varphi,
\\
\sA{}^{\hat{0}} {}_{\hat{1} 3} = \sA{}^{\hat{1}} {}_{\hat{0} 3} =\sA{}^{\hat{0}} {}_{\hat{2} 2} = \sA{}^{\hat{2}} {}_{\hat{0} 2}= -\frac{1}{2rf} \sqrt{\frac{2M}{r}} \sin\theta\sin\varphi,
\quad
\sA{}^{\hat{0}} {}_{\hat{2} 2} = \sA{}^{\hat{2}} {}_{\hat{0} 2} = \frac{1}{2rf} \sqrt{\frac{2M}{r}} \cos\theta\sin\varphi,
\nonumber\\
\sA{}^{\hat{0}} {}_{\hat{3} 1} = \sA{}^{\hat{3}} {}_{\hat{0} 1} =- \frac{1}{2rf} \sqrt{\frac{2M}{r}} \cos\theta,
\quad
\sA{}^{\hat{0}} {}_{\hat{3} 2} = \sA{}^{\hat{3}} {}_{\hat{0} 2} = -
\frac{1}{2rf} \sqrt{\frac{2M}{r}} \sin\theta,
\nonumber
\\
\sA{}^{\hat{1}} {}_{\hat{2} 3} = - \sA{}^{\hat{2}} {}_{\hat{1} 3} = \left(1- \frac{1}{f}\right)\sin^2\theta,
\quad
\sA{}^{\hat{1}} {}_{\hat{3} 2} =- \sA{}^{\hat{3}} {}_{\hat{1} 2} = - \left(1- \frac{1}{f}\right)\cos\varphi,
\label{h_A_Lem_spin}\\
\sA{}^{\hat{1}} {}_{\hat{3} 3} = - \sA{}^{\hat{3}} {}_{\hat{1} 3} =
\left(1- \frac{1}{f}\right)\sin\theta\cos\theta\sin\varphi,
\quad
\sA{}^{\hat{2}} {}_{\hat{3} 2} = - \sA{}^{\hat{3}} {}_{\hat{2} 2} = -
\left(1- \frac{1}{f}\right)\sin\varphi,
\nonumber\\
\sA{}^{\hat{2}} {}_{\hat{3} 3} = - \sA{}^{\hat{3}} {}_{\hat{2} 3} =  -\left(1- \frac{1}{f}\right)\sin\theta\cos\theta\cos\varphi. \nonumber
\end{gather}
which we can again recognize to be almost identical to \eqref{ISC_Sch_d}, but the components with $\hat{0}$ have a minus sign.

\subsection{The $e$-gauge}
We use here the shorthand notation (\ref{A_e}) and (\ref{B_e}) for ${\cal A}_e$ and ${\cal B}_e$, respectively.

1) To the tetrad $\stackrel{B}{h}$ corresponds an inertial spin connection with the following non-zero components:
\begin{gather}
\sA{}^{\hat{0}} {}_{\hat{1} 1} = \sA{}^{\hat{1}} {}_{\hat{0} 1} = -\frac{e M}{f r^2}
\l(e^2-f\r)^{-\frac{1}{2}},
\qquad
\sA{}^{\hat{0}} {}_{\hat{3} 3} = \sin\theta\sA{}^{\hat{0}} {}_{\hat{2} 2} = \sA{}^{\hat{3}} {}_{\hat{0} 3} = \sin\theta\sA{}^{\hat{2}} {}_{\hat{0} 2} = {\cal B}_e f^{-\frac{1}{2}} \sin\theta
\nonumber \\
\sA{}^{\hat{1}} {}_{\hat{3} 3} = \sin\theta \sA{}^{\hat{1}} {}_{\hat{2} 2} = - \sA{}^{\hat{3}} {}_{\hat{1} 3} = -  \sin\theta\sA{}^{\hat{2}} {}_{\hat{1} 2} = - {\cal A}_e f^{-\frac{1}{2}} \sin\theta,
\qquad
\sA{}^{\hat{2}} {}_{\hat{3} 3} = - \sA{}^{\hat{3}} {}_{\hat{2} 3} =  -\cos\theta.\label{FFspincinhB}
\end{gather}

\newpage
2) To the tetrad $\stackrel{A }{h}$ corresponds an inertial spin connection with the following non-zero components:
\begin{gather}
\sA{}^{\hat{0}} {}_{\hat{1} 1} = \sA{}^{\hat{1}} {}_{\hat{0} 1} = -\frac{e M}{f r^2}
\l(e^2-f\r)^{-\frac{1}{2}}\sin\theta \cos\varphi,
\qquad
\sA{}^{\hat{0}} {}_{\hat{2} 1} = \sA{}^{\hat{2}} {}_{\hat{0} 1} =   -\frac{e M}{f r^2}\l(e^2-f\r)^{-\frac{1}{2}}\sin\theta\sin\varphi,\nonumber
\\
\sA{}^{\hat{0}} {}_{\hat{3} 1} = \sA{}^{\hat{3}} {}_{\hat{0} 1} =   -\frac{e M}{f r^2}\l(e^2-f\r)^{-\frac{1}{2}}\cos\theta,
\qquad
\sA{}^{\hat{0}} {}_{\hat{1} 2} = \sA{}^{\hat{1}} {}_{\hat{0} 2} =
{\cal B}_e f^{-\frac{1}{2}}\cos\theta\cos\varphi,
\nonumber\\
\sA{}^{\hat{0}} {}_{\hat{1} 3} = \sA{}^{\hat{1}} {}_{\hat{0} 3} = -
{\cal B}_e f^{-\frac{1}{2}}\sin\theta\sin\varphi,
\qquad
\sA{}^{\hat{0}} {}_{\hat{2} 2} = \sA{}^{\hat{2}} {}_{\hat{0} 2} = {\cal B}_e f^{-\frac{1}{2}}\cos\theta\sin\varphi,
\label{FFspincinhA}\\
\qquad
\sA{}^{\hat{0}} {}_{\hat{2} 3} = \sA{}^{\hat{2}} {}_{\hat{0} 3} = {\cal B}_e f^{-\frac{1}{2}}\sin\theta\cos\varphi,\qquad
\sA{}^{\hat{0}} {}_{\hat{3} 2} = \sA{}^{\hat{3}} {}_{\hat{0} 2} = -{\cal B}_e f^{-\frac{1}{2}}\sin\theta,
\nonumber
\\
\sA{}^{\hat{1}} {}_{\hat{2} 3} =- \sA{}^{\hat{2}} {}_{\hat{1} 3} =
  - \left({\cal A}_e f^{-\frac{1}{2}}-1 \right)\sin^2 \theta,
\qquad
\sA{}^{\hat{1}} {}_{\hat{3} 2} = - \sA{}^{\hat{3}} {}_{\hat{1} 2} = \left({\cal A}_e f^{-\frac{1}{2}}-1 \right)\cos \varphi,
\nonumber\\
\sA{}^{\hat{2}} {}_{\hat{3} 2} = - \sA{}^{\hat{3}} {}_{\hat{2} 2} =\left({\cal A}_e f^{-\frac{1}{2}}-1 \right)\sin \varphi,
\qquad
\sA{}^{\hat{1}} {}_{\hat{3} 3} = - \sA{}^{\hat{3}} {}_{\hat{1} 3} =
-\left({\cal A}_e f^{-\frac{1}{2}}-1 \right)\sin\theta\cos\theta \sin\varphi,
\nonumber\\
\sA{}^{\hat{2}} {}_{\hat{3} 3} = - \sA{}^{\hat{3}} {}_{\hat{2} 3} = \left({\cal A}_e f^{-\frac{1}{2}}-1 \right)\sin\theta\cos\theta \cos\varphi.\nonumber
\end{gather}

\section{Triviality of $\bar{\beta}=0$ free-falling tetrad in $f(T)$ gravity\label{apptriv}}
\setcounter{equation}{0}
In section~\ref{secfT}, we have found that the antisymmetric field equations for the tetrad \eqref{tetfTD}, which was constructed in analogy with the Lemaitre case \eqref{LemproperinSch} as a Lorentz boosted  free-falling tetrad in the general spherically symmetric spacetime, are given by \eqref{ftast0}-\eqref{ftast3}. These antisymmetric field equations are satisfied only in three cases: $f_{TT}=0$, $\sT'=0$, and $\bar{\beta}=0$. While the first case restrict the theory to TEGR identically, the second case reduces to TEGR with an effective cosmological constant. The last case of $\bar{\beta}=0$ implies $A=1$, but in principle leaves us with an option of having a nontrivial $B$. Let us show that this not the case, and the remaining symmetric field equations enforce $B=1$ as well, and hence make this case trivial.

We consider a tetrad \eqref{tetfTD}, find the antisymmetric field equations \eqref{ftast0}-\eqref{ftast3}, and consider the case $\bar{\beta}=1$ that implies $A=1$. We then find that the remaining field equations in their mixed form to be
\begin{eqnarray}
E_0{}^0&=& -\frac{4(f_T+f_{TT}\sT'r)B^2-4(f_T+f_{TT}\sT'r)B+fr^2B^3+4f_T B'r}{2r^2B^3},\\
E_1{}^1&=& -\frac{f}{2}+\frac{2f_T}{r^2B^2}-\frac{2f_T}{r^2B},\\
E_2{}^2&=&-\frac{2(2f_T+f_{TT}\sT'r)B^2-2(f_T+f_{TT}\sT'r)B+(-2f_T+fr^2)B^3+2f_TB'r}{2r^2B^3}, \\
E_3{}^3&=&E_2{}^2. 
\end{eqnarray}  
We can then subtract the first and third equations to find
\begin{equation}
2E_2{}^2-E_0{}^0= - \frac{f}{2}+\frac{2f_T}{r^2} - \frac{2f_T}{r^2B}.
\end{equation}
Comparing this with the second equation we see that we must have 
\begin{equation}
B=1.
\end{equation}
Therefore, the case $\bar{\beta}=0$ is indeed trivial and there is no ``hidden" non-trivial solution. Naturally we could have guessed this right away since $\bar{\beta}=0$ corresponds to no boost at all and is just a special case of the good tetrad \eqref{tetfTprop} with $A=1$. However, we are not aware of demonstrating the triviality of $A=1$ solution in the $f(T)$ literature.

\bibliography{references}
\bibliographystyle{Style}

\end{document}